\documentclass[twocolumn]{aastex631}
\usepackage{float}
\usepackage{amsmath}
\usepackage[shortlabels]{enumitem}
\usepackage[encapsulated]{CJK}

\shorttitle{UNCOVER Morphology Catalog}
\shortauthors{Zhang et al.}

\graphicspath{{./}{figures/}}

\begin{document}

\title{Everything Every Band All at Once I: A Global Morphology Catalog in Abell 2744 based on UNCOVER/MegaScience}

\author[0000-0001-6454-1699]{Yunchong Zhang} 
\affiliation{Department of Physics and Astronomy and PITT PACC, University of Pittsburgh, Pittsburgh, PA 15260, USA}
\email{yuz369@pitt.edu}

\author[0000-0001-8367-6265]{Tim B. Miller}
\affiliation{Center for Interdisciplinary Exploration and Research in Astrophysics (CIERA), Northwestern University, 1800 Sherman Ave, Evanston, IL 60201, USA}

\author[0000-0002-0108-4176]{Sedona H. Price}
\affiliation{Space Telescope Science Institute, 3700 San Martin Drive, Baltimore, Maryland 21218, USA}

\author[0000-0002-1714-1905]{Katherine A. Suess}
\affiliation{Department for Astrophysical \& Planetary Science, University of Colorado, Boulder, CO 80309, USA}

\author[0000-0001-5063-8254]{Rachel Bezanson}
\affiliation{Department of Physics and Astronomy and PITT PACC, University of Pittsburgh, Pittsburgh, PA 15260, USA}

\author[0000-0003-4075-7393]{David J. Setton}\thanks{Brinson Prize Fellow}
\affiliation{Department of Astrophysical Sciences, Princeton University, 4 Ivy Lane, Princeton, NJ 08544, USA}

\author[0000-0001-6755-1315]{Joel Leja}
\affiliation{Department of Astronomy \& Astrophysics, The Pennsylvania State University, University Park, PA 16802, USA}
\affiliation{Institute for Computational \& Data Sciences, The Pennsylvania State University, University Park, PA 16802, USA}
\affiliation{Institute for Gravitation and the Cosmos, The Pennsylvania State University, University Park, PA 16802, USA}

\author[0000-0001-7160-3632]{Katherine E. Whitaker}
\affiliation{Department of Astronomy, University of Massachusetts, Amherst, MA 01003, USA}
\affiliation{Cosmic Dawn Center (DAWN), Denmark} 

\author[0000-0002-5612-3427]{Jenny~E.~Greene}
\affiliation{Department of Astrophysical Sciences, Princeton University, 4 Ivy Lane, Princeton, NJ 08544, USA}

\author[0000-0002-1109-1919]{Robert Feldmann}
\affiliation{Department of Astrophysics, University of Zurich, Zurich, CH-8057, Switzerland}

\author[0000-0001-7201-5066]{Seiji Fujimoto}
\affiliation{David A. Dunlap Department of Astronomy and Astrophysics, University of Toronto, 50 St. George Street, Toronto, Ontario, M5S 3H4, Canada}
\affiliation{Dunlap Institute for Astronomy and Astrophysics, 50 St. George Street, Toronto, Ontario, M5S 3H4, Canada}

\author[0000-0003-2804-0648 ]{Themiya Nanayakkara}
\affiliation{Centre for Astrophysics and Supercomputing, Swinburne University of Technology, PO Box 218, Hawthorn, VIC 3122, Australia}

\author[0000-0003-2680-005X]{Gabriel Brammer}
\affiliation{Cosmic Dawn Center (DAWN), Denmark} 
\affiliation{Niels Bohr Institute, University of Copenhagen, Jagtvej 128, K{\o}benhavn N, DK-2200, Denmark}

\author[0000-0002-7031-2865]{Sam E. Cutler}
\affiliation{Department of Astronomy, University of Massachusetts, Amherst, MA 01003, USA}

\author[0000-0001-8460-1564]{Pratika Dayal}
\affiliation{Canadian Institute for Theoretical Astrophysics, 60 St George St, University of Toronto, Toronto, ON M5S 3H8, Canada}
\affiliation{David A. Dunlap Department of Astronomy and Astrophysics, University of Toronto, 50 St. George Street, Toronto, Ontario, M5S 3H4, Canada}
\affiliation{Department of Physics, 60 St George St, University of Toronto, Toronto, ON M5S 3H8, Canada
}

\author[0000-0002-2380-9801]{Anna de Graaff}
\thanks{Clay Fellow}
\affiliation{Max-Planck-Institut f\"ur Astronomie, K\"onigstuhl 17, D-69117, Heidelberg, Germany}
\affiliation{Center for Astrophysics | Harvard \& Smithsonian, 60 Garden St., Cambridge MA 02138 USA}

\author[0000-0001-7440-8832]{Yoshinobu Fudamoto} 
\affiliation{Center for Frontier Science, Chiba University, 1-33 Yayoi-cho, Inage-ku, Chiba 263-8522, Japan}

\author[0000-0001-6278-032X]{Lukas J. Furtak}\affiliation{Cosmic Frontier Center, The University of Texas at Austin, Austin, TX 78712, USA}
\affiliation{Department of Astronomy, The University of Texas at Austin, Austin, TX 78712, USA}

\author[0000-0003-4700-663X]{Andy D. Goulding}
\affiliation{Department of Astrophysical Sciences, Princeton University, 4 Ivy Lane, Princeton, NJ 08544, USA}

\author[0000-0002-3475-7648]{Gourav Khullar}
\affiliation{Department of Astronomy, University of Washington, Physics-Astronomy Building, Box 351580, Seattle, WA 98195-1700, USA}
\affiliation{eScience Institute, University of Washington, Physics-Astronomy Building, Box 351580, Seattle, WA 98195-1700, USA}

\author[0000-0002-2057-5376]{Ivo Labbe}
\affiliation{Centre for Astrophysics and Supercomputing, Swinburne University of Technology, Melbourne, VIC 3122, Australia}

\author[0000-0002-5337-5856]{Brian Lorenz}
\affiliation{Department of Astronomy, University of California, Berkeley, CA 94720, USA}

\author[0000-0001-9002-3502]{Danilo Marchesini}
\affiliation{Department of Physics and Astronomy, Tufts University, 574 Boston Ave., Medford, MA 02155, USA}

\author[0000-0002-9816-9300]{Abby Mintz}
\affiliation{Department of Astrophysical Sciences, Princeton University, 4 Ivy Lane, Princeton, NJ 08544, USA}

\author[0000-0002-8530-9765]{Lamiya A. Mowla}
\affiliation{Dunlap Institute for Astronomy and Astrophysics, 50 St. George Street, Toronto, Ontario, M5S 3H4, Canada}

\author[0000-0002-9330-9108]{Adam Muzzin}
\affiliation{Department of Physics and Astronomy, York University, 4700 Keele St., Toronto, Ontario, M3J 1P3, Canada}

\author[0000-0002-7524-374X]{Erica J. Nelson}
\affiliation{Department for Astrophysical \& Planetary Science, University of Colorado, Boulder, CO 80309, USA}

\author[0000-0002-9651-5716]{Richard Pan}
\affiliation{Department of Physics and Astronomy, Tufts University, 574 Boston Ave., Medford, MA 02155, USA}

\author[0009-0001-0715-7209]{Natalia Porraz Barrera}
\affiliation{Department for Astrophysical \& Planetary Science, University of Colorado, Boulder, CO 80309, USA}

\author[0000-0002-5522-9107]{Edward N. Taylor}
\affiliation{Centre for Astrophysics and Supercomputing, Swinburne University of Technology, Melbourne, VIC 3122, Australia}

\author[0000-0002-5027-0135]{Arjen van der Wel}
\affiliation{Sterrenkundig Observatorium, Universiteit Gent, Krijgslaan 281 S9, 9000 Gent, Belgium}

\author[0000-0001-9269-5046]{Bingjie Wang (\begin{CJK*}{UTF8}{gbsn}王冰洁\ignorespacesafterend\end{CJK*})}
\thanks{NHFP Hubble Fellow}
\affiliation{Department of Astrophysical Sciences, Princeton University, 4 Ivy Lane, Princeton, NJ 08544, USA}

\author[0000-0003-1614-196X]{John R. Weaver}\thanks{Brinson Prize Fellow}
\affiliation{MIT Kavli Institute for Astrophysics and Space Research, 70 Vassar Street, Cambridge, MA 02139, USA}

\author[0000-0003-2919-7495]{Christina C.\ Williams}
\affiliation{NSF’s National Optical-Infrared Astronomy Research Laboratory, 950 North Cherry Avenue, Tucson, AZ 85719, USA}


\begin{abstract}
We present spectrally-resolved structural parameter measurements of 28,274 sources from the legacy lensing field of Abell 2744, quantifying global structures from observed $0.7 \mu m - 4.8 \mu m$ and spanning rest-frame UV to NIR at $R\sim15$. These measurements are made on imaging mosaics mainly from the UNCOVER/MegaScience survey, including 20 JWST NIRCam broad and medium bands. We perform single-component S\'ersic fitting to these galaxies using \texttt{pysersic}, a Bayesian structural fitting tool, to infer their structural parameters and associated random uncertainties from the posterior distributions. Through various quality evaluation criteria, we infer robust structural parameters among $> 85\%$ of the selected $\rm SNR>10$ sources. For each galaxy with reliable sizes in at least two bands and a high quality redshift, we fit its observed size as a function of wavelength and infer rest-frame UV, optical, and near-infrared sizes where applicable. By performing injection-recovery tests on simulated galaxy cutouts in selected bands, we establish that our structural parameter measurements achieve fractional error $< 10 -20\%$ above $\rm SNR>10$. With this paper, all raw structural measurements and fitted rest-frame sizes are quality-flagged, cataloged, and released to the community. Finally, we demonstrate that this catalog enables the structural study of galaxies over an unprecedentedly wide parameter space of redshift ($0.3<z<8$), stellar mass ($\rm 10^{7}\, M_{\odot}<M_{*} <10^{11.5}\, M_{\odot}$), and rest-frame optical size ($\rm 100 \,pc<R_{e}<10\,kpc$), after correcting for lensing magnification. 

\end{abstract}

\section{Introduction} \label{sec:intro}

Key physical processes that foster or regulate the growth of galaxies throughout cosmic time, such as gas accretion onto the dark matter halo, stellar or active galactic nuclei feedback, and galaxy mergers, jointly shape the galaxy structure --- i.e., the geometric distributions of stellar populations, gas, and dust in galaxies. As these various components either emit or attenuate light, the observed morphology of a galaxy imprints information about its intrinsic 3D physical structures. Therefore, the study of galaxy morphology provides insight to help shape our understanding of galaxy evolution \citep[e.g.,][]{Conselice.2014}.

One of the most fundamental morphological properties of a galaxy is its size. A common size estimate is the half-light radius (i.e., the radius that encloses 50\% of the light), although some work has explored the benefit of other definitions \citep[e.g.,][]{Mowla.etal.2019}. Traditionally, this relies on fitting a parametric form to the 2D light profile \citep[i.e., S\'ersic profile;][]{Sersic.1963}, which can account for the point-spread function (PSF) via forward-modeling. This is especially valuable at high redshifts, where the imaging PSF can be substantial compared to the galaxy itself. 

Taking advantage of the early imaging sky surveys, such as the Sloan Digital Sky Survey \citep{York.etal.2000}, studies have shown that in the nearby universe, galaxy size follows a logarithmic normal distribution at a given magnitude or stellar mass, the median galaxy size scales with galaxy stellar mass, and this scaling relation is steeper for red elliptical galaxies than blue disks \citep[e.g.,][]{Shen.etal.2003,Lange.etal.2015}. Space-based instruments (and to a lesser extent ground-based instruments) have successfully characterized galaxy morphologies of relatively massive galaxies slightly beyond cosmic noon ($z<3$) through imaging surveys such as the Cosmic Assembly Near-infrared Deep Extragalactic Legacy Survey \citep{Grogin.etal.2011,Koekemoer.etal.2011}, the Hubble Frontier Fields \citep{Lotz.etal.2017}, and the Hyper Supreme-Cam Subaru Strategic Program \citep{Aihara.etal.2018}. These surveys have enabled the study of the redshift evolution of the mass-size scaling relationship \citep[e.g.,][]{vanderWel.etal.2014,Shibuya.etal.2015,Mowla.etal.2019,Nedkova.etal.2021,Kawinwanichakij.etal.2021,Cutler.etal.2022}. Using parametric fits, studies have also examined other structural properties, such as the surface or central stellar densities \citep[e.g.,][]{Barro.etal.2017,Whitaker.etal.2017}, galaxy mass profiles \citep[e.g.,][]{Mosleh.etal.2017,Suess.etal.2019}, and 3D shapes \citep[e.g.,][]{vanderWel.etal.2014b,Zhang.etal.2019}. These studies have established and fostered a consensus that the sizes of extended star-forming galaxies grow slowly with mass via star formation, while the sizes of more compact spheroidal quiescent galaxies grow rapidly with mass via dry merging \citep[e.g.,][]{vanDokkum.etal.2015}. While it remains controversial whether the morphological transformation is concurrent with quenching \citep{Barro.etal.2013,Williams.etal.2014}, it has been argued that reaching a central density threshold is likely the prerequisite of quenching \citep[e.g.,][]{vanDokkum.etal.2015,Barro.etal.2017,Whitaker.etal.2017, Mosleh.etal.2017,Suess.etal.2021}.

Despite the success at $z<3$, these high-resolution space-based imaging surveys lack access to wavelengths beyond observed-frame $2 \mu m$ and rely on rest-frame Ultraviolet (UV) morphology for those that study galaxy structures at $z>3$, which is dominated by the youngest stellar population and can be very susceptible to dust attenuation. It is also challenging to study the smaller ($\rm <1\,kpc$) and less massive ($\rm <10^9\,M_{\odot}$) galaxy population, limited by the resolution and sensitivity of these instruments. This landscape has been drastically reshaped since the launch of JWST \citep{Gardner.etal.2006}, leading to a surge of studies on rest-optical galaxy structures in previously inaccessible redshift regimes \citep[e.g.,][]{Ferreira.etal.2022,Kartaltepe.etal.2023,Baggen.etal.2023,Ono.etal.2023,Ito.etal.2024}. Notably, a handful of early studies have probed the high-redshift galaxy mass-size relation, finding smaller median galaxy sizes, but similar slope and scatter as in the local relation \citep[e.g.,][]{Ito.etal.2024,Ward.etal.2024,Wright.etal.2024,Allen.etal.2025,Yang.etal.2025}. JWST has also unlocked the ability to study the rest-frame near-infrared morphology of galaxies at cosmic noon~\citep{suess.etal.2022,miller.etal.2022,martorano.etal.2023,vanderwel.etal.2024,Martorano.etal.2025}. However, comprehensively characterizing the redshift evolution of the galaxy mass-size relation throughout all currently accessible epochs requires a statistical sample of galaxies with reliable size measurements over a large redshift and mass range. One can rely on the multi-band JWST imaging surveys with the best spectral resolution and deepest effective depths for such a sample, though it has not been assembled.

Some early JWST surveys target the blank legacy fields \citep{Eisenstein.etal.2023,Finkelstein.etal.2023,Donnan.etal.2024} and obtain wide-area NIRCam \citep{Rieke.etal.2023} broad-band images. These surveys yield large samples that are advantageous for studying the brightest and rarest objects \citep{Genin.etal.2025,Carreira.etal.2026,McGrath.etal.2026} but lack the depth to constrain the smaller or fainter galaxy population. Alternatively, other surveys make use of the most powerful strong gravitational lensing clusters to boost our probe of distant galaxies in the background. An outstanding example is Abell 2744, which has been targeted extensively by many early JWST programs, including an extended and contiguous imaging mosaic from the Ultradeep NIRSpec and NIRCam Observations before the Epoch of Reionization \citep[UNCOVER; ][]{Bezanson.etal.2024} and Medium-bands Mega Science \citep[MegaScience; ][]{Suess.etal.2024} programs. Beyond UNCOVER/MegaScience, Abell 2744 has also been targeted by a multitude of JWST programs. The combined imaging depth and magnification boost from gravitational lensing has allowed the community to probe the underexplored regime of intrinsically faint sources. Since the imaging release of these surveys, a handful of diverse aspects of galaxy morphology have been explored, including the sizes of dwarf quiescent galaxies \citep{Cutler.etal.2024,Cutler.etal.2025}, the mass-size relation at the epoch of re-ionization \citep{Miller.etal.2025}, the sizes of proto-cluster members \citep{Pan.etal.2025}, and the dust geometry of submillimeter-detected galaxies \citep{Price.etal.2025a}. These studies have advanced our understanding of galaxy structures in unexplored parameter regime, by selecting unique subsets of galaxies in this dataset. However, there has not been a study of galaxy structures that employs a standardized methodology to explore the entirety of this dataset, leveraging its full wavelength coverage.

In this paper, we present the morphology catalogs in Abell 2744, including the single-component S\'ersic fits of 28,274 unique Signal-to-Noise-Ratio (SNR) $\rm >10$ sources across 20 JWST/NIRCam bands from \texttt{pysersic} \citep{Pasha_and_Miller.2023} and the interpolated sizes at rest-frame wavelengths. We also accompany these results with a suite of recovery tests to quantify systematic uncertainties in these fits and to examine the robustness of \texttt{pysersic}-reported random uncertainties. We note that this paper is published in combination with a companion paper that leverages these catalogs to characterize the redshift-evolving galaxy mass-size relation \citep{Miller.etal.2026}. These catalogs are available online at \url{https://zenodo.org/records/18808251}.

The structure of this paper is as follows. In Section \ref{sec: data}, we describe the NIRCam mosaic images, PSFs, and ancillary catalogs from the UNCOVER/MegaScience survey used in this study. In Section \ref{sec: method}, we describe the single-component S\'ersic fit procedure, quality flags, and direct measurement catalogs. In Section \ref{sec: recovery test}, we detail the generation and fitting of simulated source images and present the analysis of systematic and random uncertainties in our fits. In Section \ref{sec: rest-frame size}, we demonstrate the methodology of inferring rest-frame sizes and showcase the rest-frame optical size versus stellar mass in various redshift bins. Finally, we provide of brief summary of this study in Section \ref{sec: summary}. Throughout this paper, we assume a flat $\mathrm{\Lambda CDM}$ cosmology with $ \mathrm{\Omega_{\Lambda} = 0.71}$, $ \mathrm{\Omega_{m} = 0.29}$, and $\mathrm{H_{0} = 69.32 \, km\,s^{-1}\,Mpc^{-1}}$ as reported in \cite{Hinshaw.etal.2013}.

\section{Data} \label{sec: data}
\subsection{JWST Imaging Mosaics of Abell 2744}

The morphology catalog presented in this study is based on JWST/NIRCam image mosaics targeting the lensing cluster A2744 at $z = 0.307$, primarily from the UNCOVER \citep[GO-2561;][]{Bezanson.etal.2024} and MegaScience \citep[GO-4111;][]{Suess.etal.2024} surveys. The image mosaics analyzed in this work also contain observations from the following programs: MAGNIF (GO-2883, PI: Sun), ALT (GO-3516, PI: Naidu \& Matthee; \citealp{Naidu.etal.2024}), GO-3538 (PI: Iani), ERS-1324 (GLASS, PI: Treu; \citealp{Treu.etal.2022}), and DD-2756 (PI: Chen; \citealp{Chen.etal.2022}). The raw JWST data corresponding to these mosaics were obtained from the Mikulski Archive for Space Telescopes (MAST) at the Space Telescope Science Institute \footnote{The specific original observations analyzed can be accessed via doi: \dataset[10.17909/7yvw-xn77]{http://dx.doi.org/10.17909/7yvw-xn77}.}. These mosaics span the entire NIRCam filter suite, reaching $\rm 28-30 \,mag \,AB$ in depth. An area of $\sim \rm 30 \, arcmin^2$ is covered by all filters. 

The specific science and weight image mosaics analyzed in this work correspond to UNCOVER Data Release 3. The method used to produce these mosaics is detailed in \cite{Weaver.etal.2024}. In brief, these mosaics are reduced and drizzled to a common grid of 0.02 arcsec/pixel at short wavelengths (shorter than F250M; SW) or 0.04 arcsec/pixel at long wavelengths (F250M and longer; LW). In the science image mosaics, the foreground bright cluster galaxies (bCGs) and intra-cluster light (ICL) are modeled and subtracted. In addition, we obtain the corresponding exposure time maps that are grid-matched to these science image mosaics from the DAWN JWST Archive (DJA)\footnote{These exposure time maps can be accessed at \href{https://dawn-cph.github.io/dja/imaging/v7/}{https://dawn-cph.github.io/dja/imaging/v7/}.}. We also make use of the PSFs empirically constructed from the stars in these mosaics in each band, which are introduced in \cite{Weaver.etal.2024}.

\subsection{Ancillary Catalogs}
In this study, we rely on the UNCOVER/MegaScience Data Release 3 (DR3) photometric catalog to select targets for structural fitting and to set up fitting priors. This catalog was originally presented in \cite{Weaver.etal.2024}, and updated to include the medium band imaging from MegaScience in \cite{Suess.etal.2024}. In brief, the source detection in this catalog was performed on a noise-equalized F277W+F356W+F444W stacked image, and the aperture photometry extraction was conducted on a PSF-matched image mosaic for each band, using \texttt{aperpy}\footnote{Available at \href{https://github.com/astrowhit/aperpy}{https://github.com/astrowhit/aperpy}.}. We use the segmentation map produced from the source detection process to produce image masks where needed. 

To infer physical sizes at rest-frame wavelengths, we use the redshift ($z_{50}$) and source magnification reported in the UNCOVER Data Release 4 (DR4) stellar population synthesis (\texttt{SPS}) catalog \citep{Wang.etal.2024}, which was updated to make use of the spectroscopic redshifts and v2.0 lens model in \cite{Price.etal.2025b}\footnote{The UNCOVER DR3 photometric catalog and DR4 \texttt{SPS} catalog are available at \href{https://jwst-uncover.github.io/}{https://jwst-uncover.github.io/}.}. In brief, the stellar population properties in this catalog were derived from spectral energy distribution (SED) modeling of the photometry from the \texttt{SUPER} photometric catalog \citep{Weaver.etal.2024}, utilizing \texttt{Prospector} \citep{Leja.etal.2017,Johnson.etal.2021} with informative, redshift-dependent \texttt{Prospector-{$\beta$}} priors \citep{Wang.etal.2023}. Using the source magnification map given by the Abell 2744 lens model \citep{Furtak.etal.2023}, properties such as stellar mass and redshift are estimated simultaneously for self-consistent solutions. In $\sim 800$ cases where a spectroscopic redshift is available, the redshift is fixed to the spectroscopic redshift \citep{Price.etal.2025b}. We also use the rest-frame colors and stellar masses presented in the \texttt{SPS} catalog for data visualization in Section \ref{sec: rest-frame size}.

\section{Structural Fitting} \label{sec: method}
\subsection{Target Selection and Structural Fitting Setup} \label{sec: fit setup}

To start, we take the UNCOVER/MegaScience DR3 photometric catalog and iterate through all catalog entries in all JWST NIRCam bands. We select any source in any given band to perform single-S\'ersic fitting if it satisfies the following criteria:

\begin{itemize}
\item{${ use\_phot} = 1$;}
\item{$\rm SNR > 10$, where $\rm SNR \equiv f_{\nu,band}/e_{\nu,band}$.}
\end{itemize}
Our choice of SNR-based selection aims to exclude sources where we expect large fractional error and to reduce overall computation cost. We demonstrate the correlation between SNR and fractional error with recovery tests in Section \ref{sec: recovery test}. Our selection results in an initial sample that contains from $\sim 6000$ to $\sim 27000$ targets in each of the 20 NIRCam bands, with the specific number for each band presented in Table \ref{tab:N fit}. There are typically more eligible ($\rm SNR>10$) targets in broad bands than medium bands, as the broad band images reach deeper effective depths \citep{Weaver.etal.2024,Suess.etal.2024}.

\begingroup
\setlength{\tabcolsep}{16pt}
\begin{table*}[t!]
\caption{ This table summarizes the relevant statistics of our fits in each filter in this study. We fit all $\rm SNR>10$ sources in each filter with the \texttt{SVI-MVN} method and all $\rm 10<SNR<100$ sources with the \texttt{MCMC} method. We show these statistics separately for the fits performed with the \texttt{SVI-MVN} or \texttt{MCMC} method in the second and third broad columns. In each broad column, we sequentially show the number of selected targets, the number of fits evaluated as \texttt{use\_fit=2}, and the number of fits evaluated as \texttt{use\_fit=1}, in a given filter. We also show the fraction of total fits evaluated as \texttt{use\_fit=2} or \texttt{use\_fit=1}. When the fraction of sources evaluated as the given quality is negligible (i.e., less than 0.1\%), we show a hyphen instead. The \texttt{use\_fit} quality flag is detailed in Section \ref{sec: flags}. }
    \centering
    \begin{tabular}{c|ccc|ccc}
    \hline
    \hline
      Filter & &\texttt{SVI-MVN} & & &\texttt{MCMC}& \\
       & & $\rm SNR>10$ & & &$\rm 10<SNR<100$& \\
    &$\rm N_{fit}$& $\rm N_{use\_fit=2} $ &$\rm N_{use\_fit=1} $&$\rm N_{fit}$&$\rm N_{use\_fit=2} $&$\rm N_{use\_fit=1} $\\
     \hline
     F070W & 6915 & 5747 (83.1\%)& 74 (1.1\%)&6145&3686 (60.0\%) & 2 (-)\\
     F090W & 16484 & 13867 (84.1\%) & 310 (1.9\%)&14547& 8827 (60.7\%) & 13 (0.1\%)\\
     F115W & 17259 & 15465 (89.6\%) & 322 (1.9\%) & 15165& 9977 (65.8\%) & 12 (0.1\%) \\
     F140M & 4685 & 4051 (86.5\%) & 48 (1.0\%) & 4122 &2687 (65.2\%) &1 (-) \\
     F150W & 18938 & 16732 (88.4\%) & 370 (2.0\%) & 16515 &10728 (65.0\%)& 18 (0.1\%) \\
     F162M & 5937 & 5120 (86.2\%) & 57 (1.0\%) & 5095 &3232 (63.4\%) &1 (-) \\
     F182M & 8191 & 6812 (83.2\%) & 108 (1.3\%) &7042 &4202 (59.7\%) &1 (-) \\
     F200W & 23398 & 20403 (87.2\%) & 465 (2.0\%) & 20255 &12610 (62.3\%) &36 (0.2\%)\\
     F210M & 7558 & 6407 (84.8\%) & 100 (1.3\%) & 6460 & 3997 (61.9\%) & 1 (-)\\
     F250M & 5845 & 5262 (90.0\%) & 166 (2.8\%) & 4969 & 3236 (65.1\%) & 11 (0.2\%)\\
     F277W & 26007 & 21574 (83.0\%) & 1119 (4.3\%)& 22500 & 12815 (57.0\%) &  87 (0.4\%)\\
     F300M & 8171 & 7367 (90.2\%) & 263 (3.2\%) & 7047 & 4501 (63.9\%) & 22 (0.3\%)\\
     F335M & 7717 & 6979 (90.4\%) & 242 (3.1\%) &6612 & 4168 (63.0\%) & 12 (0.2\%)\\
     F356W & 26552 & 22356 (84.2\%) & 1113 (4.2\%) &22840 & 12834 (56.2\%) & 63 (0.3\%)\\
     F360M & 7456 & 6727 (90.2\%) & 273 (3.7\%) & 6359 &4030 (63.4\%) &50 (0.7\%)\\
     F410M & 9196 & 8383 (91.2\%) & 276 (3.0\%) & 7854 & 4935 (62.8\%) & 6 (0.1\%)\\
     F430M & 3996 & 3714 (92.9\%) & 75 (1.9\%) & 3263 & 2184 (66.9\%)& 0 (-) \\
     F444W & 19103 & 16391 (85.8\%) & 893 (4.7\%) & 16553 & 9727 (58.8\%) & 37 (0.2\%)\\
     F460M & 3433 & 3176 (92.5\%) & 60 (1.7\%) & 2787 & 1887 (67.7\%) & 1 (-) \\
     F480M & 4023 & 3739 (92.9\%) & 58 (1.4\%) & 3255 & 2150 (66.1\%) & 1 (-) \\
      \hline
    \end{tabular}
    \label{tab:N fit}
\end{table*}
\endgroup

For each target, our pipeline prepares a suite of grid-matched cutouts, including a science image, an error image, and a mask, prior to the implementation of structural fitting. Starting with the science image, we adopt a square cutout centered on the flux center of the main target from the detection image. We initialize the cutout image with a minimum width of $4''$, and we then iteratively increase the cutout width by 5 pixels in all directions. At every step after enlarging the cutout, we check the distance between the cutout edge and the nearest segment edge of the target, using the segmentation map. When the source segment edge is at least $\rm 0.5''$ from the cutout edge, we stop enlarging the cutout and finalize the cutout width. A maximum width of $12''$ is enforced for these cutouts. The half of the maximum cutout width is larger than the Kron radii of all sources in UNCOVER/MegaScience, except for one galaxy, which has a $\rm R_{eff} \sim 1.2''$ in both SW and LW bands. Therefore, our selected maximum cutout width is sufficiently large for fitting all sources in this dataset. We extract the corresponding cutouts from the error map and the segmentation map, adopting the same centroid location and the width as the science cutout. 

The error map $\rm err(x,y)$ is produced using the science image $\rm sci(x,y)$, exposure time map $\rm exp(x,y)$, and the weight image $\rm wht(x,y)$ as follows:
\begin{equation}
\rm err(x,y) = \sqrt{1/wht(x,y) + \sigma_{poisson}(x,y)},
\end{equation}
where $\rm \sigma_{poisson}(x,y)$ is the Poisson variance map, defined as:
\begin{equation}
\rm \sigma_{poisson}(x,y) = \frac{max(sci(x,y),0)}{(exp(x,y)\cdot \texttt{phot\_scale})},
\end{equation}
where the \texttt{phot\_scale} is a multiplicative factor that describes the net effective gain of the \texttt{grizli} produced science image\footnote{Following the methodology detailed in the DJA tutorials at \href{https://dawn-cph.github.io/dja/blog/2023/07/18/image-data-products/}{https://dawn-cph.github.io/dja/blog/2023/07/18/image-data-products/}.}.

When the primary target is located close to bright extended neighboring sources, the contamination of the neighboring flux can lead to an overestimation of the flux attributed to the primary target and bias the inference of its structural parameters. To mitigate such biases, we choose to simultaneously model any additional sources in the science cutout with a single S\'ersic model, if these sources:
\begin{itemize}
\item{are no more than 2 magnitude fainter than the primary target in F356W, F277W, and the band in which the fit is performed;}
\item{have $\rm SNR>5$ in F356W, F277W, and the band in which the fit is performed\footnote{We choose to omit the first two requirements in other bands, since the median area-weighted effective depth of the UNCOVER/MegaScience image mosaic is mostly much shallower in other bands than in F356W or F277W \citep{Weaver.etal.2024}.};}
\item{are less than $\rm 2 ''$ from the centroid position of the primary target from the detection image.}
\end{itemize}
Sources that do not meet the above criteria are masked in the fitting process. We produce masks for each cutout suite by taking the source regions from the segmentation map and dilating them to be 3 pixels larger with a tophat filter.

\subsection{Implementation with \texttt{pysersic}}\label{sec: implementation}

We perform single-S\'ersic fitting of all selected primary targets and neighboring sources in each cutout using  \texttt{pysersic}~\citep{Pasha_and_Miller.2023}, which is a Python-based Bayesian inference framework for structural fitting accelerated with \texttt{Jax} \citep{jax2018github}. Compared to traditional structural fitting tools such as \texttt{GALFIT} \citep{Peng.etal.2002,GALFIT} that rely on chi-square minimization, \texttt{pysersic} provides more robust uncertainties on the inferred structural parameters by estimating the posterior distribution of these parameters with Bayes' theorem. It has been demonstrated that there is no significant systematic difference between the best-fitting results obtained by these two methods \citep{Zhang.etal.2024}, though \texttt{GALFIT} tends to severely underestimate the associated uncertainties \citep{Haussler.etal.2007,van.der.Wel.2012,Cutler.etal.2022}.

For each cutout, we subtract a flat sky background whose amplitude is fixed by the median value of the unmasked pixels. Each single-S\'ersic model has seven free parameters. The x and y centroid positions are in pixels and have Gaussian priors centered on the source positions reported in the photometric catalog, with a half-width of 0.5 pixels. For flux, we assume a truncated Gaussian prior that is centered on the flux reported in the photometric catalog, has a width ($\sigma$) taken to be half of the photometric catalog flux, and has a minimum floor of 0. For the half-light radius, we assume a log-uniform prior whose lower limit is always set to be 0.25 pixels. Meanwhile, its upper limit is set to be 40, 70, or 100 pixels, if the flux radius in pixels ($\rm R_{flux}$) reported in the photometric catalog falls in the following range: $[0,30)$, $\rm [30,60)$, or $\rm [60,\infty)$, respectively. We inflate the upper limits and associated $\rm R_{flux}$ ranges by a factor of two when fitting SW bands, where the pixel scale is a factor of two smaller than LW bands. The ceiling of 100 pixels (or 200 pixels in SW bands) is larger than the flux radii of almost all sources in UNCOVER/MegaScience, except for $\sim300$ sources. We visually inspected the model residuals of these sources. None of them are physically large sources with $R_{eff}$ greater than 100 pixels ($\rm 4''$), and their flux radii are typically significantly overestimated due to failure in deblending. The S\'ersic index, position angle, and ellipticity have uniform priors between $[0.65,8]$, $\rm [0,2\pi]$, or $\rm [0,0.9]$, respectively. The details of these prior choices are summarized in Table \ref{tab:priors}.

For all primary targets, we first estimate the Maximum A Posteriori (MAP) values for all free parameters, using the built-in \texttt{find\_MAP} method of \texttt{pysersic}. We do not report these point estimates in the final catalogs, but we use them to evaluate the quality of posterior estimation, which is described in Section \ref{sec: flags}. We then perform a fiducial estimation of the S\'ersic parameter posterior through the \texttt{SVI-MVN} (stochastic variational inference using a multivariate normal) method. SVI is a method of approximating the posterior by assuming a parametric form of the distribution \citep{wingate.etal.2013,Kucukelbir.etal.2017}. The parameters of the chosen distribution family are then optimized by finding the ``best fit'' to the posterior, often achieved by minimizing the KL divergence~\citep{kullback.etal.1951}. This frames inference as an optimization problem and is much faster than traditional sampling techniques at the cost of only being an approximation, i.e. limited by the flexibility of the chosen parameterization. For S\'ersic profile fitting, we assume multivariate Gaussian and optimize the means and covariances of each parameter to find the best fit to the posterior. 

For high SNR sources where the posteriors are tight and well behaved, we find that the \texttt{SVI-MVN} method works well and provides reasonable estimates of the uncertainty in a much shorter time. However, low SNR sources may have more complex posteriors. For any sources with $\rm 10<SNR<100$, we opt to perform an additional posterior estimation with traditional \texttt{MCMC} (Markov chain Monte Carlo) sampling, using the No U-turn Sampler (NUTS; \citealp{Hoffman.etal.2011,Phan.etal.2019}). In each instance sampled with \texttt{MCMC}, we run two chains for 500 warm-up iterations and 1000 sampling steps. We expect the \texttt{MCMC} sampling to provide more accurate uncertainties, especially in the case of non-Gaussian posterior distributions. The choice of the SNR cutoff ($\rm SNR<100$) is motivated by the fact that random uncertainty of structural parameters can be typically constrained down to $5\%$ above this limit (see Section \ref{sec: recovery test} or the similar analysis presented in \cite{van.der.Wel.2012}). The total uncertainty budget for $\rm SNR>100$ sources is more likely to be dominated by systematic uncertainties that our fitting procedure cannot account for. In this case, mapping the precise posterior distribution with \texttt{MCMC} is not necessary or particularly meaningful, given its higher computational cost\footnote{\texttt{MCMC} typically requires at least $300\%$ more computational time per instance than \texttt{SVI-MVN}, given the setup in this work.}.

\begingroup
\setlength{\tabcolsep}{10pt}
\begin{table*}[t!]
\caption{The priors of free parameters in each single S\'ersic model used in this study.}
    \centering
    \begin{tabular}{ccc}
    \hline
    \hline
    Parameter & Description & Prior\\
     \hline
    sky &  Amplitude of the uniform sky background   &  Fixed to median value of unmasked pixels\\
     \hline
    xc  & S\'ersic centroid in x-axis (pixels) & Gaussian: $\mu = x_{c, phot}$, $\sigma = 0.5$ \\
    yc  & S\'ersic centroid in y-axis (pixels) & Gaussian: $\mu = y_{c, phot}$, $\sigma = 0.5$ \\
    $\rm f_{\nu}$ & Integrated flux of the S\'ersic model ($\rm 10nJy$) & Truncated Gaussian: $\mu = f_{\nu, phot}$, $\sigma = f_{\nu, phot}/2$, min = 0\\
    n  & S\'ersic index & Uniform: $\rm min=0.65$,  $\rm max=8$ \\
    $\rm R_{eff}$ & Half-light Radius (pixels) & Log Uniform: $\rm min=0.25$,  $\rm max=40$, if $R_{flux} <30$;\\
    & & Log Uniform: $\rm min=0.25$,  $\rm max=70$ , if $30<R_{flux} <60$;\\
    & & Log Uniform: $\rm min=0.25$,  $\rm max=100$ , if $60<R_{flux}$\\
   $\rm \epsilon $  & Ellipticity & Uniform: $\rm min=0$,  $\rm max=0.9$ \\
    $\rm \theta$  & Angular position in radian & Uniform: $\rm min=0$,  $\rm max=2\pi$ \\
      \hline
    \end{tabular}
    \tablecomments{($x_{c, phot}$, $y_{c, phot}$), $f_{\nu, phot}$, and $R_{flux}$ refer to the source centroid in x and y axis, source flux, and flux radius of the corresponding target in the photometric catalog. Ellipticity ($\rm \epsilon $) is defined as one minus the ratio of the semi-minor axis over the semi-major axis ($1-b/a$). Angular position ($\rm \theta$) is defined as the clockwise angle between the semi-major axis and the positive y direction. The three-tier prior on $R_{eff}$ listed above corresponds to fits in LW imaging. In SW imaging, where the pixel scale is a factor of two smaller, the upper bound of $R_{eff}$ prior in each tier and the associated $R_{flux}$ bounds in the conditioning are all a factor of two larger.}
    \label{tab:priors}
\end{table*}
\endgroup

\subsection{Quality Checks and Flags} \label{sec: flags}

We determine the quality of every fit by checking the following criteria:

\begin{enumerate}[A)]
\item The number of masked pixels is less than 50\% of the total number of pixels enclosed by four times the best-fitting half-light radius of the primary target.
\end{enumerate}
This removes instances where the primary target model is driven to over-extrapolate masked data.

\begin{enumerate}[A), resume]
\item The MAP value of half-light radius, S\'ersic index, and flux are within a reasonable range from the 50th percentile posterior values. We choose this range explicitly as: 
\begin{equation}
\rm log\left(\frac{P_{MAP}-P_{50th}}{(P_{84th}-P_{16th})/2}\right) <1, 
\end{equation}
where $\rm P_{MAP}, P_{16th}, P_{50th},P_{84th}$ denote the MAP value, 16th, 50th, and 84th percentile posterior values of each parameter, respectively. 
\end{enumerate}
This removes instances where the posterior is trapped in a regional maxima in the likelihood space that mismatches the global maxima indicated by the MAP value.

\begin{enumerate}[A), resume]
\item The half-light radius posterior is not pushing against the upper bound ($\rm R_{eff,84th} \in [39,40], [69,70], or \,[99,100]$) while the S\'ersic index posterior is not pushing against the lower bound ($\rm n_{16th} \in [0.65, 0.75]$). 
\end{enumerate}
This eliminates a particular failure mode in which the model is driven to be unrealistically flat when the primary target is faint and extended or when there are artifacts in the background. This failure mode typically occurs when the primary target is faint and diffuse, and the model is dominated by sky flux rather than the flux from the primary target.
\begin{enumerate}[A), resume]
\item The posterior chain converges. We compute the minimum bulk effective sample size ($\rm ESS_{b,min}$), minimum tail effective sample size ($\rm ESS_{t,min}$), and maximum $\hat{r}$ convergence diagnostic ($\rm \hat{r}_{max}$) of all chains. We define the convergence criteria as $\rm ESS_{b,min}>250$, $\rm ESS_{t,min}<250$, and $\rm \hat{r}_{max}<1.05$.
\end{enumerate}
We note that criterion E) only applies to \texttt{MCMC} fits.
\begin{enumerate}[A), resume]
\item The chi-squared of the unmasked residual per pixel is less than 2. Specifically, we define this value as: 
\begin{equation}
    \rm \chi_{pp}^2 = \frac{1}{N_{pixel}}\sum \frac{(sci(x,y)-mod(x,y))^2}{err(x,y)^2}.
\end{equation}

\end{enumerate}
Such a flux-dependent cutoff evaluates whether the observed data is well-described by the best-fitting model via chi-squared per pixel statistics. Sources with non-single-S\'ersic characteristics, such as spiral arms or irregular clumps, may fail this requirement. Using the $\rm \chi_{pp}^2$ metric, we classify sources passing through all other criteria into two quality categories that will be detailed later.

\begin{enumerate}[A), resume]
\item The inferred 50th percentile centroid positions in x and y have to be within $0.12''$ (3 pixels for LW bands; 6 pixels for SW bands) from the source centroid positions in the photometric catalog. 
\end{enumerate}
The specific value of this cutoff roughly corresponds to $8\sigma$ of the distribution in centroid offset for this sample, which approximately follows a 2D Gaussian distribution. This is a conservative choice that eliminates instances whose best-fitting centroid position extremely deviates from the flux center in the photometric catalog. This type of failure occurs when the primary target is faint while bright neighbors are present. The single S\'ersic model is consequently dominated by the flux gradient of neighboring sources instead of the primary target, drifting away from the photometric catalog centroid.

\begin{enumerate}[A), resume]
\item The difference between the inferred median value for flux and the photometric catalog value has to be less than 2 magnitudes. 
\end{enumerate}
This eliminates instances whose flux significantly deviates from that in the photometric catalog. This type of issue often occurs concurrently when the fit also fails Criterion C) and/or Criterion F). In addition, in cases where the source has a distinct bulge-disk composition, the single S\'ersic component may be driven by either the bulge or the disk, and consequently yield an inconsistent flux estimate relative to the photometric catalog value. 

\begin{enumerate}[A), resume]
\item The inferred 50th percentile half-light radius is not 3 times larger than the semi-major-axis of the flux (i.e, second-order moment of the detection image; ``a\_image'') from the photometric catalog. 
\end{enumerate}
This removes unphysically large fits when the model is dominated by sky flux or contaminated by nearby bright neighbors, similar to the failure mode removed by Criterion C).

\begin{enumerate}[A), resume]
\item The surface brightness of the S\'ersic profile at the 50th percentile half-light radius is larger than the value corresponding to the median $3\sigma$ image depth in the given band. 
\end{enumerate}
This similarly removes unphysically large fits that are dominated by sky flux, by requiring sufficiently high SNR in the outskirts of S\'ersic fits.

However, dropping out fits according to these quality evaluation criteria complicates the overall selection bias and increases the sample incompleteness, in addition to our $\rm SNR>10$ target selection. We will examine and discuss the completeness of these fits as a function of stellar mass and redshift in Appendix \ref{appendix:A}.

\begin{figure*}[!htb]
    \centering
    \includegraphics[width = 1.\textwidth]{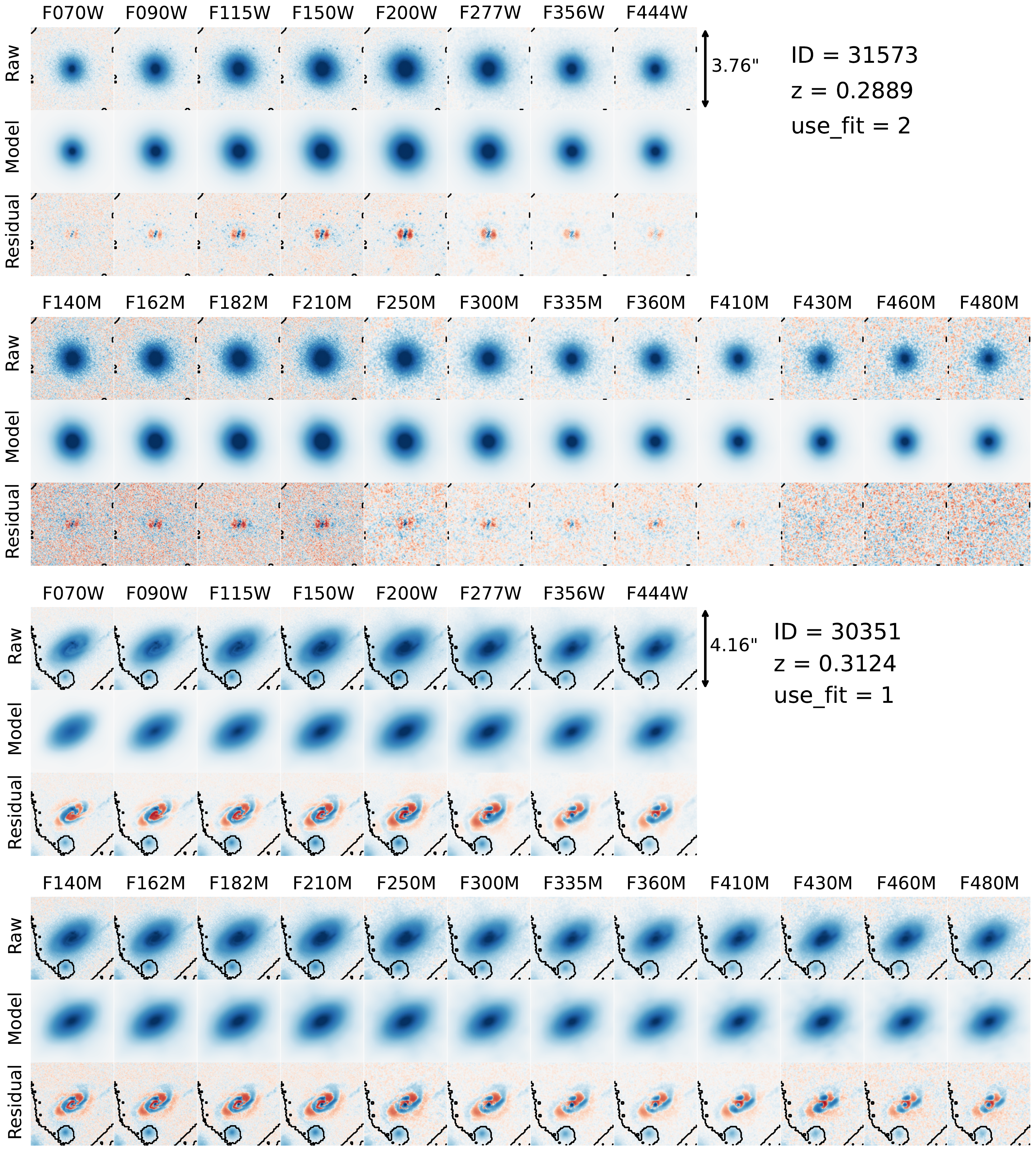}
    \vspace{20pt}
    \caption{Example S\'ersic fits of two galaxies in all 20 NIRCam medium or broad bands. The six rows on the top show the original images, best-fitting models, and residuals of a galaxy (ID: 31573) that is well-described by a single 2D S\'ersic model (use fit = 2). The six rows on the bottom present the fitting results of a galaxy (ID: 30351) that has non-S\'ersic structural features, such as spiral arms, but reliable size measurements (use fit = 1). The black solid contours show the image mask used in structural fitting. These fits are properly flagged and allow us to characterize galaxy structural parameters as functions of wavelength.}
    \label{fig:img_mod_residual}
\end{figure*}

We institute a three-level classification metric for labeling the quality of these fits. The fits deemed to have great quality (equivalent to \texttt{use\_fit = 2} in the \texttt{RAW} catalog, introduced in Section \ref{sec: catalogs}) have to meet all of the above criteria. Sources in this category are well-described by a single-S\'ersic model in the given band, and all of their best-fitting parameters are deemed robust. A selected example of a galaxy in this category (in all 20 bands) is shown in the top rows (ID: 31573) in Figure \ref{fig:img_mod_residual}. 

Robust fits (equivalent to \texttt{use\_fit = 1} in the \texttt{RAW} catalog) meet all the above criteria, but fail Criteria E) due to high chi-squared per pixel ($\rm \chi_{pp}^2>2$) values. Sources in this category are not well-approximated by a single-S\'ersic model. In this case, there are often asymmetries that lead to a high chi-squared value. However, the S\'ersic profile still reasonably represents the overall light profile. A selected example of a galaxy in this category is shown in the bottom rows (ID: 30351) in Figure \ref{fig:img_mod_residual}. However, we further enforce a different cutoff in $\rm \chi_{pp}^2$ for fits in this category, in order to eliminate fits where the asymmetries are so significant that the reported structural parameters are not meaningful. Specifically, we require: 
\begin{equation}
\rm \chi_{pp}^2 <4,
\end{equation}
if $\rm log(f_{\nu}/10nJy) <1$, or
\begin{equation}
\rm log(\chi_{pp}^2) < (log(f_{\nu}/10nJy)-1)\cdot{0.5} + log(4),
\end{equation}
if $\rm log(f_{\nu}/10nJy) >1$. Here, $f_{\nu}$ denotes the photometric catalog flux of the target source in the fitting band. After visually inspecting various randomly selected instances with $\rm \chi_{pp}^2$, we choose this cutoff to be more tolerant for brighter sources, where we expect asymmetries to be more prominent. We also require any fits classified in this category to be resolved (i.e, $\rm R_{eff} > 1 $ pixel). We generally encourage catalog users to trust the sizes in this category, but to be mindful of particular use cases when adopting other best-fitting parameters.

Finally, we classify any sources that fail at least one of the above criteria (except for Criteria E)) as failed instances, which is equivalent to \texttt{use\_fit = 0} in the \texttt{RAW} catalog. We don't recommend users to adopt any parameters of sources in this category. The number of fits attempted and success rate (\texttt{use\_fit = 1 or 2}) in each band by each method are summarized in Table \ref{tab:N fit}.

For each source, we additionally provide a \texttt{flag\_nearbcg} and a \texttt{mu\_flag} that may benefit the usage of these S\'ersic parameters. The \texttt{flag\_nearbcg} concerns the potential pollution of any unsubtracted bCG light in these fits if these sources are close to any bCGs. A source has \texttt{flag\_nearbcg = 1} if it is within $3''$ of any known bCGs. We migrated this flag from the DR3 photometric catalog \citep{Weaver.etal.2024}. The \texttt{mu\_flag} concerns the significant distortion of the source image when its magnification due to cluster gravitational lensing is high. In addition, any underlying complex structures, such as clumps, can be enhanced by high lensing magnification \citep{Fujimoto.etal.2025,Nakane.etal.2025,Vanzella.etal.2025} and therefore may not be well modeled by an ideal S\'ersic profile. We assign a source \texttt{mu\_flag = 0} if the source has an average magnification $\rm \mu<4$ ($\sim 98\%$ of all sources in the \texttt{RAW} catalog) and \texttt{mu\_flag = 1} above $\rm \mu>4$.

We note that the centroid positions of best-fitting S\'ersic models have a median sub-pixel (0.005'' - 0.01'') offset from their photometric catalog source centroids. These offsets vary in amplitude and direction across different bands, and they are insignificant relative to the mosaic pixel scale ($\sim $ a quarter of a pixel). These offsets likely originate from the sub-pixel misalignment between the intrinsic centers of the empirically-constructed PSF and the PSF pixel grid. We document these offsets and discuss their correlation with the sub-pixel misalignment in PSFs in Appendix \ref{appendix:B}.

\subsection{Catalogs}\label{sec: catalogs}

\begin{figure}[t!]
    \centering
    \includegraphics[width = 1.\linewidth]{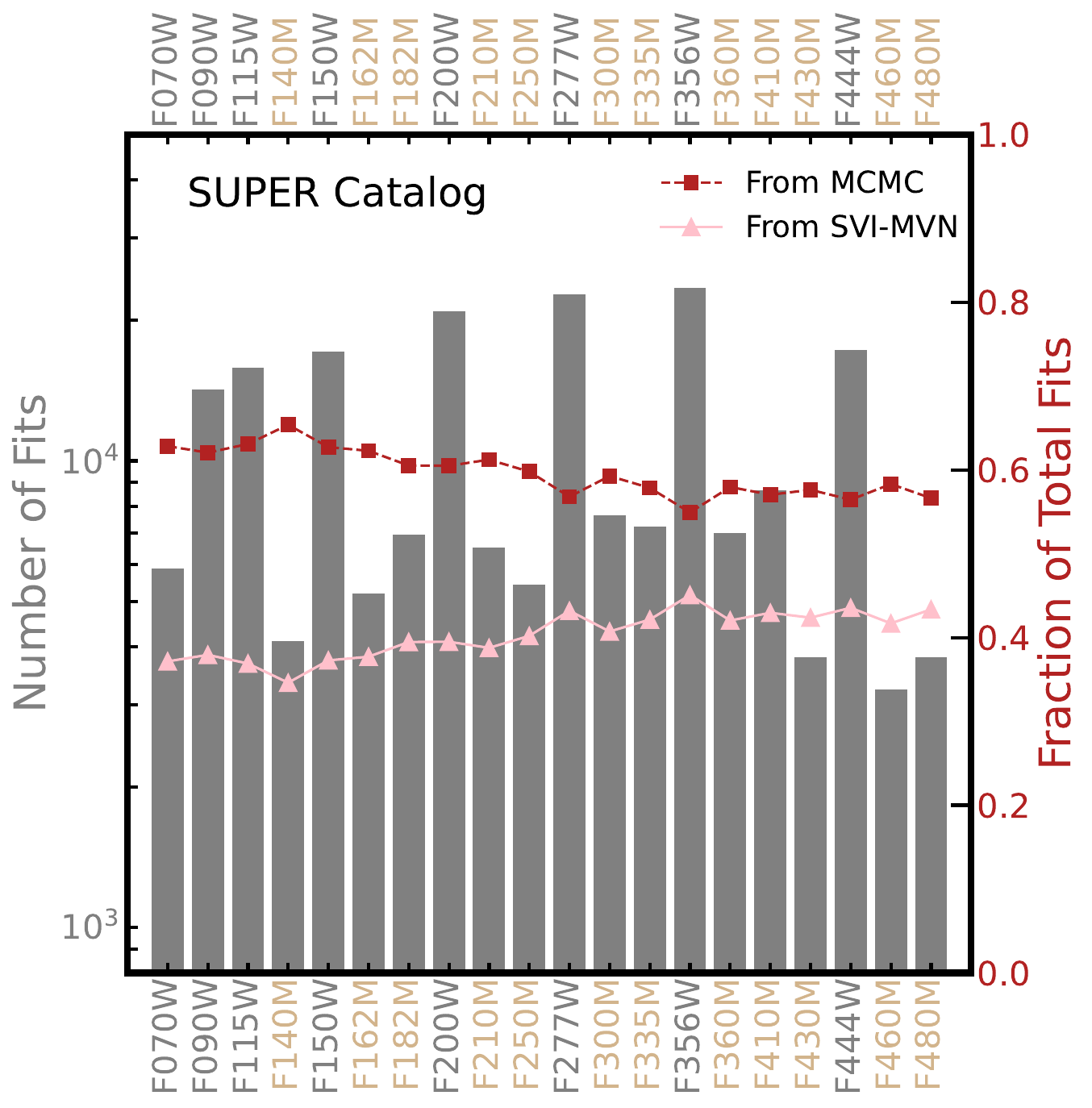}
    \caption{ The gray histogram shows the distribution of number of available fits in the \texttt{SUPER} catalog over available bands (left axis). The red or pink curves show the fraction of those fits obtained from either the SVI-MVN or MCMC method as functions of all 20 NIRCam medium or broad bands (right axis). Roughly $60\%$ of these fits are retrieved from full MCMC sampling, which better characterizes the parameter posterior. The total number of available fits is around 10 to 20 thousand in most of the deep broadband images. In $\rm \sim 1\, mag$ shallower medium band images, the total number of successful fits drops to 4 to 7 thousand. Overall, 28,274 unique galaxies have acceptable fits in at least one band by either one of the \texttt{pysersic} methods.}
    \label{fig:morph stat}
\end{figure}

In the \texttt{RAW} catalog, we record the 16th/50th/84th posterior percentile of all S\'ersic parameters, all statistics involved in fit quality evaluation, and fit quality flags of all attempted S\'ersic fits in all bands, regardless of posterior estimation method (\texttt{MCMC} versus \texttt{SVI-MVN}) or fit quality. This catalog is row-matched to the UNCOVER/MegaScience DR3 photometric catalog, with which it also shares the same ID convention. We summarize the number and the percentage fraction (over attempted fits) of fits labeled as great or robust quality (\texttt{use\_fit = 1 or 2}) in each band in Table \ref{tab:N fit}. We list all columns in the \texttt{RAW} catalog and detail their descriptions in Appendix \ref{appendix:C}. The \texttt{RAW} catalog serves as a reservoir for all original outputs of our structural fitting pipeline. It becomes useful when the customized or complex usage of these structural fits is needed (i.e., customizing quality flags; searching for emission line galaxies by checking [even failed] fits in particular bands). 

\begin{figure*}[!htb]
    \centering
    \includegraphics[width = 1.\textwidth]{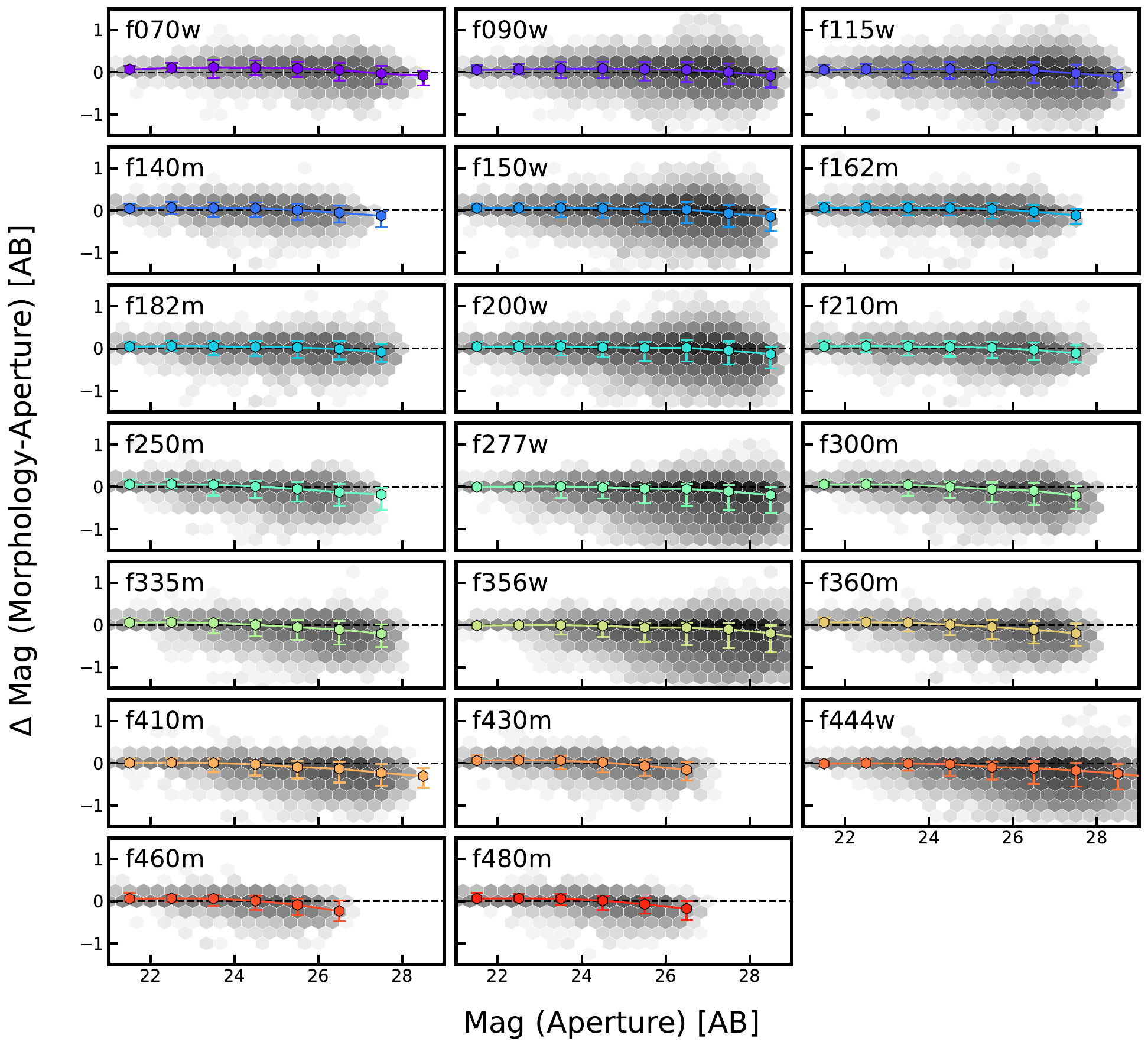}
    \caption{The difference in magnitude from this work (structural fitting) and the photometry catalog (aperture photometry) as a function of photometry catalog magnitude, for all galaxies in the \texttt{SUPER\_DUPER} catalog in all 20 NIRCam medium or broad bands. We also show the population median and $1\sigma$ scatter in magnitude bins as error bars. Based on the median of these distributions, we find no significant systematic differences between the photometry catalog fluxes and \texttt{pysersic} recovered fluxes at $\rm m_{AB} <26$. However, structural fitting reports systematically higher fluxes than aperture photometry at $\rm m_{AB} >26$, especially in the broadband images, where sources to a fainter limit are selected for fitting. These systematic differences in the faint regime are likely driven by the intrinsic difference between the S\'ersic profile and the aperture loss correction, in terms of the assumption of the surface brightness curve of growth in the source outskirt.}
    \label{fig:flux_vs_flux}
\end{figure*}

However, the \texttt{RAW} catalog has $\sim 1000$ columns, and it is inflated by information that might not be relevant in most scientific use cases, such as failed fits or different versions of parameter posterior estimated by either method. Therefore, we produce a \texttt{SUPER} catalog that serves as a ``go-to" option by filtering raw results and exclusively including only the best quality fits for each source. For the \texttt{SUPER} catalog, we iterate through every source in each band and report the parameter posterior percentiles obtained from \texttt{MCMC} as long as they have \texttt{use\_fit = 1 or 2}. If \texttt{MCMC} results have \texttt{use\_fit = 0}, we instead report these results obtained from \texttt{SVI-MVN} as long as they have \texttt{use\_fit = 1 or 2}. If both \texttt{MCMC} and \texttt{SVI-MVN} results are labeled with \texttt{use\_fit = 0}, we leave the source entry blank (filled with \texttt{NaN} value). We opt to only report posterior percentiles for flux, half-light radius, S\'ersic index, and ellipticity. Instead of reporting posterior percentiles of centroid position in the x or y pixel-frame axis, we report the distance between the best-fitting S\'ersic centroid and the source centroid reported in the photometric catalog, in arcseconds. We note that centroids in two given bands with the same offset distance may not be exactly co-spatial, as these distances we report merely represent the offset amplitude while omitting the direction. The \texttt{SUPER} catalog inherits the ID, RA, and Dec columns from the photometric catalog and shares the same \texttt{flag\_nearbcg} and \texttt{mu\_flag} as the \texttt{RAW} catalog. We list all columns in the \texttt{SUPER} catalog and detail their descriptions in Appendix \ref{appendix:C}. In Figure \ref{fig:morph stat}, we show the number and fractional composition (obtained through \texttt{MCMC} or \texttt{SVI-MVN}) of available fits per band in the \texttt{SUPER} catalog. The overall large number of available fits is secured by the full coverage at $\rm SNR>10$ and high success rate ($\sim 85\%$, see Table \ref{tab:N fit}) of the \texttt{SVI-MVN} method. In all bands, most ($\sim 60\%$) of the available fits come from the \texttt{MCMC} method, which provides more robust estimation of the parameter posteriors when they are complex and non-Gaussian.

We similarly produce a more aggressively filtered \texttt{SUPER} catalog where we only adopt results with \texttt{use\_fit = 2}, the \texttt{SUPER\_DUPER} catalog, which only contains the structural parameters of sources that are well-described by a single S\'ersic profile. In Figure \ref{fig:flux_vs_flux}, we examine the differences in magnitudes retrieved from fitting S\'ersic models in \texttt{SUPER\_DUPER} catalog to photometric catalog magnitudes that were aperture-extracted. In all 20 bands, there are no significant systematic discrepancies between the S\'ersic model fluxes and the aperture-extracted fluxes at $m_{AB}<26$. However, for fainter sources, a S\'ersic model systematically infers higher flux (or lower magnitude) value than an aperture-corrected extraction. This trend likely reflects the intrinsic difference in accounting for fluxes in the source outskirts between these methodologies. It is uncertain whether either of these methods suffers from systematic errors in estimating flux in the faint regime. We further explore systematic uncertainties in S\'ersic models in the next section (Section \ref{sec: recovery test}). We stress that the structural measurements reported in the \texttt{RAW}, \texttt{SUPER}, and \texttt{SUPER\_DUPER} catalogs are all image-plane quantities (i.e, not corrected for magnification due to gravitational lensing). Our choice of retaining the original measurements allows users to flexibly switch to any other magnification corrections than the default one from the UNCOVER v2.0 lens model \citep{Furtak.etal.2023, Price.etal.2025b}, in case the lens model is improved in the future.

In this work, we also produce a ``Rest Frame Size Catalog" where we report inferred galaxy sizes at given rest-frame wavelengths, which we will describe in details in Section \ref{sec: rest-frame size}.

\section{Quantifying the Uncertainties with Recovery Testing}\label{sec: recovery test}
Understanding the systematic uncertainties in these inferred S\'ersic parameters is critical but challenging, since there is no ground truth for comparison. Inspired by the procedure in \cite{van.der.Wel.2012}, we implement a suite of recovery tests where we generate simulated images by injecting mock galaxies (assuming single S\'ersic profiles) into real empty sky cutouts from the mosaics, perform the same structural fitting routine, and compare the ground truth to inferred parameters. This experiment is designed to characterize the systematic uncertainties in S\'ersic parameters as a function of SNR (or, comparably, magnitude), given realistic degrading effects due to PSF and sky background. However, it does not characterize the systematic effects due to any errors in the PSF models or any intrinsic non-S\'ersic features in these sources, since mock galaxies are generated as ideal S\'ersic profiles convolved with the same PSFs used in fitting. We perform these tests in three bands: F200W, F250M, and F444W. These bands represent three scenarios: an SW broad band (20 mas/pixel resolution), a medium band (shallower effective depth than broad bands), and an LW broad band (40 mas/pixel resolution).

\subsection{Creating Mock Galaxies}

For each of these three bands, we extract 100 square cutouts that are 4'' wide (101 pixel times 101 pixel for an LW band; 201 pixel times 201 pixel for an SW band) from empty regions in the mosaic, which are identified using the segmentation map. However, we are unable to sample the sky background in the core region of the cluster, which is extremely crowded. We also require these cutouts to be extracted from mosaic regions with similar exposure time.

For each band, we generate a suite of 1000 mock galaxies that are PSF-convolved single S\'ersic profiles. 600 of these mock galaxies are designed to be ``bright" and trace the realistic source population with magnitudes between 21 and 27 in the given band. We choose 27 as the cutoff magnitude for this bright population because roughly all sources have $\rm SNR>10$ above this cutoff. We choose S\'ersic parameters by randomly drawing the flux, half-light radius, S\'ersic index, and ellipticity (in the same band) of sources in the \texttt{SUPER\_DUPER} catalog within this magnitude range. We add random noise to these ``true'' parameters, aiming to mimic rather than completely duplicate the catalog parameter distributions. The random noise is drawn from Gaussian distributions that are centered at zero and have widths arbitrarily chosen to be $\sim 1\%$ of the parameter value. 

One of the motivations behind these recovery tests is to explore the limit in magnitude (or SNR) at which these structural fits become significantly unreliable due to large systematic or random uncertainties. Therefore, a ``faint" population of mock galaxies is needed to sample the S\'ersic parameter distributions in the low SNR regime. However, the S\'ersic parameter distributions below our SNR cut ($\rm SNR>10$) are not included in our existing catalogs. We instead choose to create another 400 mock galaxies as possible analogs of the ``faint" source population. Similar to creating the bright population, we randomly draw noise-added S\'ersic parameters from realistic sources with magnitudes between 26 and 27. We then randomly assign magnitudes between 27 and 30, adopting the faint cutoff magnitude roughly comparable to the deepest $\rm 5\sigma$ depth in these images. 

\begin{figure*}[!ht]
    \centering
    \includegraphics[width = 1.\textwidth]{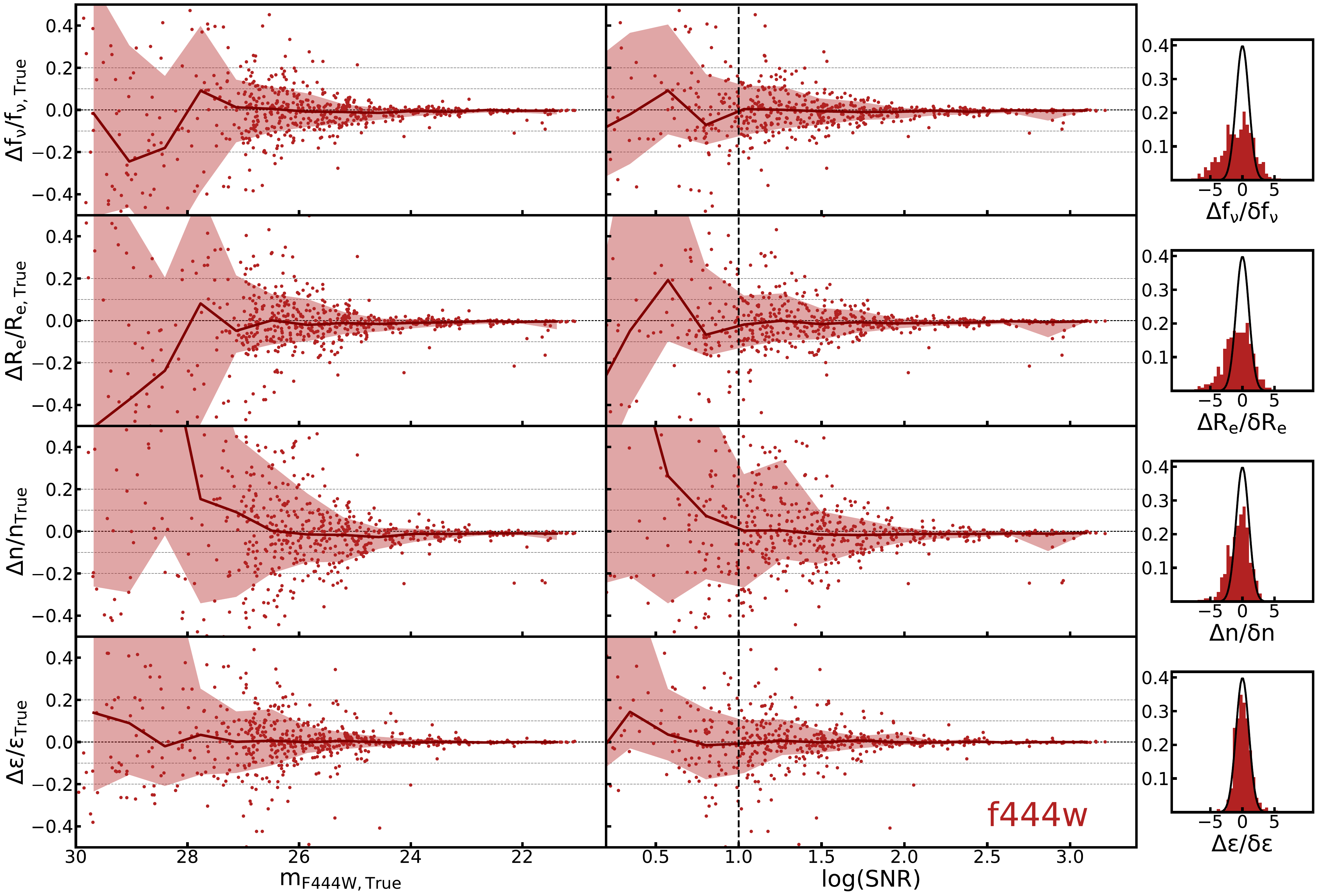}
    \caption{Recovery tests in F444W: The S\'ersic parameter fractional error (fitted value minus the true value, normalized by the true value) versus true magnitude (first column from the left) or SNR (second column from the left). From top to bottom, each row shows the fractional error on flux, half-light radius, S\'ersic index, and ellipticity. The accuracy of recovered S\'ersic parameters decreases with source brightness or SNR. In F444W, \texttt{pysersic} retrieves S\'ersic parameters within fractional errors of 10\% for the majority ($68\%$) of the instances at a given magnitude or SNR, down to $\rm SNR\sim 30$, which approximately corresponds to $\rm m_{F444W} \sim 26$. There is little ($<2\%$) or no systematic offset in the median fractional errors at $\rm SNR>10$ (or $\rm m_{F444W} < 27$). The panels on the right show the histogram of S\'ersic parameter residuals normalized by \texttt{pysersic} uncertainties at $\rm SNR>10$, compared to a unit Gaussian distribution. Notably, \texttt{pysersic} uncertainties encapsulate the discrepancy between true and fitted values exceptionally well for ellipticity. However, \texttt{pysersic} systematically underestimates the uncertainty for flux, half-light radius, and S\'ersic index by a factor of roughly 2 (or 1.5, in the case of S\'ersic index), as indicated by broadening of these distributions relative to the unit Gaussian distribution.}
    \label{fig:recovery_f444w}
\end{figure*}

For each mock galaxy in each band, we randomly select an empty sky cutout and inject the mock galaxy at the cutout center. Because the modeling priors depend on aperture photometry, we measure the SNR, aperture flux, and flux radius of each mock instance, following the method in \cite{Weaver.etal.2024}. We've also prepared the error image for these mock instances, using the same procedure described in Section \ref{sec: fit setup}. We neglect the mask image, since these mock galaxies are designed as isolated sources. We defer testing the performance of \texttt{pysersic} in crowded fields to future studies. For each mock instance, we set S\'ersic parameter priors with these measured aperture fluxes and flux radii and sample the parameter posterior with \texttt{MCMC}. The exact prior choices and \texttt{MCMC} chain setup are identical to those in Section \ref{sec: implementation}. Finally, we evaluate the quality of these mock galaxy fits, following Section \ref{sec: flags}.

\subsection{Evaluating Robustness of Fits and Uncertainties}

In Figure \ref{fig:recovery_f444w}, we show the fractional error (difference between \texttt{pysersic} 50th percentile value and the ground truth, normalized by the ground truth) in flux, half-light radius, S\'ersic index, and ellipticity as functions of magnitude or SNR for mock galaxies in F444W. We only include fits with \texttt{use\_fit = 2}, bin them by magnitude or SNR, and indicate median (solid red line) and 16th to 84th percentile scatter (light red region) fractional errors. We also show horizontal lines equivalent to no error, 10\%, and 20\% for reference. We note that the majority of these fits ($80-90\%$) pass the quality evaluation criteria at $\rm SNR>10$. As indicated by the median fractional error curves, there is little ($<2\%$) or almost no systematic error among the \texttt{pysersic}-retrieved 50th percentile parameter values, if the source has $\rm SNR>10$ (approximately equivalent to $\rm m_{F444W} < 27$ in UNCOVER). Taken from the $1\sigma$ scatter, these parameters are within $\sim$10\% for the majority of the instances at a given SNR (or magnitude), at approximately $\rm SNR > 30$ (or $\rm m_{F444W} < 26$). Below $\rm SNR \sim 6$ (or $\rm m_{F444W} \sim 28$), these inferences become unreliable. More than $60\%$ of the instances in this regime fail the quality evaluation criteria, and the remaining successful instances systematically have large fractional errors ($> 20\%$). This cutoff in SNR is similar to the one shown by the test performed with HST images \citep{van.der.Wel.2012}.

In the right panels, we plot the density distribution of error (difference between \texttt{pysersic} 50th percentile value and the ground truth, denoted by the $\rm \Delta$ prefix) over uncertainty (width of the \texttt{pysersic} $\rm 1 \sigma$ posterior, denoted by the $\rm \delta$ prefix). We only include the instances with $\rm SNR > 10$ in these distributions. Also shown is the curve of a unit Gaussian distribution. We expect these distributions to be broader than that of a unit Gaussian if \texttt{pysersic} uncertainties are underestimated relative to the true error, and vice versa. Overall, these distributions for flux, half-light radius, or S\'ersic index are broader than a unit Gaussian distribution. This suggests that the \texttt{pysersic}-reported uncertainties for these parameters are underestimated. These distributions also slightly skew towards the negative direction, indicating that parameters are significantly underestimated for a subset of galaxies. Similar skewness in recovering half-light radius at $\rm SNR\sim 5-10$ has been reported in \cite{Miller.etal.2025}. For ellipticity, the \texttt{pysersic}-reported uncertainties almost perfectly represent the true errors, as its error over uncertainty distribution closely follows that of a unit Gaussian. Overall, we find that the \texttt{pysersic} uncertainties should be inflated by a factor of 2 (for flux or half-light radius) or 1.5 (for S\'ersic index) in order for these error-over-uncertainty distributions to converge to a unit Gaussian. We opt to report the original uncertainties inferred by \texttt{pysersic} in the \texttt{RAW}, \texttt{SUPER}, and \texttt{SUPER\_DUPER} catalogs, though we generally recommend users to inflate the uncertainties of these parameters as above.

\begin{figure}[t!]
    \centering
    \includegraphics[width = 1.\linewidth]{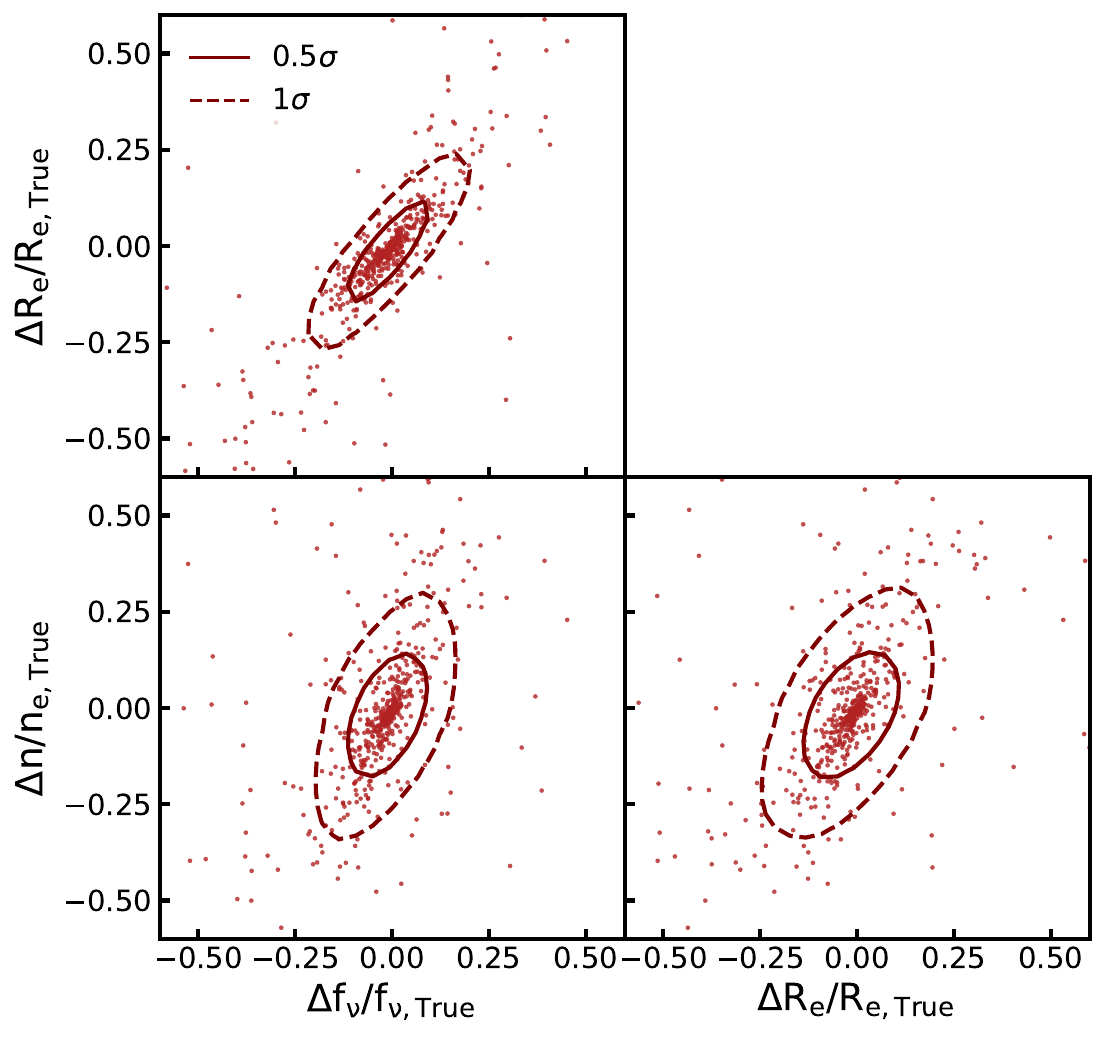}
    \caption{The corner plot showing the scatter of fractional errors in flux, half-light radius, and S\'ersic index. For any mock instance where the flux is overestimated, its half-light radius (or S\'ersic index) is more likely to be overestimated, and vice versa. }
    \label{fig:sersic degeneracy}
\end{figure}

The minor but similar skewnesses in the inferred flux, half-light radius, and S\'ersic index are due to the covariance amongst these parameters \citep{van.der.Wel.2012}. It is interesting to characterize this covariance. In Figure \ref{fig:sersic degeneracy}, we show the distribution of the fractional errors in these parameters versus each other. We overlay the contours illustrating the $0.5 \sigma$ and $1\sigma$ scatter. It is clear that there is a positive covariance among the flux, half-light radius, and S\'ersic index. Whenever the flux for a source is overestimated, the model preferentially chooses an overestimated half-light radius (and S\'ersic index), and vice versa. To quantify the correlations between fractional errors in these S\'ersic parameters, we calculate the Pearson correlation coefficient \citep{Pearson1895}, where a score of 1 corresponds to a perfect positive correlation, a score of -1 corresponds to a perfect negative correlation, and a score of 0 suggests no correlation. We restrict the calculation for pairs with fractional errors less than 60\% to avoid extreme outliers and find Pearson correlation coefficients of 0.44 between flux and S\'ersic index (p-value $\sim 10^{-27})$, 0.78 between flux (p-value $\sim 10^{-120})$ and half-light radius, and 0.49 between S\'ersic index and half-light radius (p-value $\sim 10^{-33})$. This is a known behavior due to the parameterization of the S\'ersic profile. For a S\'ersic profile, the central surface brightness ($\rm < 1Re$) remains similar when the S\'ersic index and half-light radius both become slightly larger. In contrast, a larger S\'ersic index and a larger half-light radius elevate the surface brightness in the outskirts ($\rm \gg 1Re$) of the profile, which also results in a higher integrated flux. Since the SNR in the source outskirt is typically too low to robustly differentiate between these two scenarios, these parameters appear to be covariant in these inferences, especially for faint sources. 

We have also observed similar systematic behaviors in mock tests performed in other bands, which we defer to Appendix \ref{appendix:D}. In summary, since we only select sources with $\rm SNR>10$ for fitting S\'ersic profiles in this study, we expect the flux, half-light radius, and S\'ersic index in our fits to be robust, with only marginal systematic underestimation ($<2\%$). It is shown that our reported \texttt{pysersic} uncertainties can be underestimated by a factor of 2 for flux and half-light radius (or a factor of 1.5 for S\'ersic index). We note that the underestimation of parameter uncertainties can be further exacerbated in real sources by various potential systematic uncertainties, including but not limited to the real galaxies not being ideal S\'ersic profiles, not accounting for the uncertainty in the PSF, and non-uniform sky background. In addition, cosmological dimming of surface brightness impacts the outskirts and centers of galaxies differently, which may contribute to the systematic uncertainty budget when fitting surface brightness profiles \citep{Sun.etal.2024}.

\section{Parameterizing Size as a Function of Wavelength}
\label{sec: rest-frame size}

One of the strengths of our catalog is that it covers the full 20 bands of NIRCam. To enable easier interpretation of these multi-wavelength sizes, we parameterize the size of each galaxy as a function of wavelength. This facilitates high-precision measurements by combining information from multiple nearby bands and provides a framework to discuss wavelength-dependent morphology. For this catalog we will parameterize the $\log R_e$ as a quadratic in $\log \lambda$ following:

\begin{equation}\label{eqn:size_lambda}
\begin{split}
\mathrm{log R_{e} (\lambda) = p_2 \,[log(\lambda/\mu m) - 0.35]^2} \\
\mathrm{+ p_1  \, [log(\lambda/\mu m) - 0.35] + p_0,}
\end{split}
\end{equation}
where $p_0$, $p_1$, and $p_2$ are free parameters fit for each galaxy. An offset is applied to $\log \lambda\sim 2.2\, \mu m $, which is roughly the average wavelength of the NIRCam bands, to normalize the wavelengths and facilitate easier fitting. We find that this parameterization provides the right level of simplicity versus flexibility to approximate the majority of all galaxies in our sample. We only use size measurements with \texttt{use\_fit}$ = 1,2$, fit in the observed frame wavelength, with $R_e$ in arcseconds, and limit the fitting to galaxies with $\geq 2$ bands with suitable measurements. To account for the likely underestimation of the $R_{\rm eff.}$ uncertainty observed in the Section \ref{sec: recovery test}, we add 3\% error in quadrature to all radii measurements\footnote{The mean fractional error in size is $\sim 3\%$ in a given band in this sample. Adding an additional 3\% in quadrature to the reported uncertainties is roughly comparable to our prescription of inflating size uncertainty by a factor of 2.}. Additionally, to account for the error in our PSF models, we add an additional term:
\begin{equation}
    \sigma_{R_{e},PSF} = \frac{\sigma_{\rm PSF}}{R_{\rm PSF}}\times \left(\frac{R_{\rm PSF}}{R_{\rm eff}}\right)^2
\end{equation}
in quadrature to the size uncertainty. $R_{\rm PSF}$ is the width of the PSF in each band. To obtain this width, we take the FWHM of the PSF and convert it to the standard deviation by dividing it by $\sqrt{8\ln2}$, assuming the PSF profile is approximately Gaussian. We assume a 10\% error on the PSF ($\frac{\sigma_{\rm PSF}}{R_{\rm PSF}} = 0.1$) to encapsulate the width difference between empirical PSFs and drizzled model PSFs \citep{Weaver.etal.2024}.

We use \texttt{numpyro}~\citep{Phan.etal.2019} to fit this model to each galaxy. In some cases, possible bad fits can make it through our quality cuts, or there are sharp transitions in the size as a function of wavelength, such as line emission in a medium band. To account for these scenarios, we add an outlier model following ~\citet{hogg.etal.2010}. For $p_0$ we use a broad prior of a normal distribution with a center of -0.75 and a width of 0.5, in units of $\log\, R_e / {\textrm{arcsec}}$. For $p_1$ and $p_2$ we use a student-t prior centered on 0, a width of 0.5, and 3 degrees of freedom. Our choice of priors encourages a diminishing quadratic term (i.e., $p_2 \sim 0$) when there are only two available bands in our fits. In this case, the functional form in these fits becomes effectively linear to avoid over-flexibility. We assume the outlier distribution to be log-uniform, same as input priors on $R_e$ (in Table \ref{tab:priors}). We fit for an outlier fraction $q$, to which we assign a truncated normal prior with a mean of 0 and a scale of 0.2, bounded between 0 and 0.5. We perform sampling with the NUTS sampler, using 4 chains with 500 warm-up and 1000 sampling steps each. To ensure proper convergence, we do not report fits with ESS $<100$ or $\hat{r} > 1.05$. 

In the Rest Frame Size catalog, we report the results of the sampling as the median of the posterior for the parameters $p_0$, $p_1$, and $p_2$ along with the 16th and 84th percentiles. We note that these parameters are often highly covariant, so we urge caution when using these values. As our preferred output, we also provide an additional file with the means and co-variance matrices for each galaxy. This is our recommended method of interacting with the results since it effectively captures the correlations present. With these values reported, it is possible to use this functional form to interpolate the radii to any desired wavelength. 

For convenience, we use these parameterizations to calculate the rest-frame sizes at three common wavelengths: UV at $0.2\ \mu m$, optical at $0.5\ \mu m$, and near-infrared (NIR) at $1\ \mu m$. We report the value for each size and uncertainties as the [16,50,84] percentile intervals. For each wavelength, we report two quantities: the angular size in arcseconds and the physical size in kpc. For the angular sizes, the uncertainty takes into account uncertainty in the photometric redshift, using posterior draws from the prospector-$\beta$ catalog~\citep{Wang.etal.2023, Suess.etal.2024} and the uncertainty in parameterized size at the given observed wavelength, using posterior draws of $p_0$, $p_1$ and $p_2$. For the reported physical sizes, the uncertainty also includes variation in angular diameter distance and magnification as a function of redshift. We ``de-lens'' these physical sizes using the simple prescription of $1/\sqrt{\mu}$, in contrast to other observed sizes reported in this work, where we opt to report raw measurements without any $\mu$ correction. This correction is valid for moderate magnifications $\mu\lesssim 5$. More magnified sources require more bespoke modeling that we do not perform. We report rest-frame sizes for every galaxy while minimizing extrapolation, including rest-frame wavelengths within 0.1 dex of the shortest and the longest wavelength band with successful size measurements. A full description of the columns in this table is included in Appendix \ref{appendix:C}.

\begin{figure*}[!htb]
    \centering
    \includegraphics[width = 1.\linewidth]{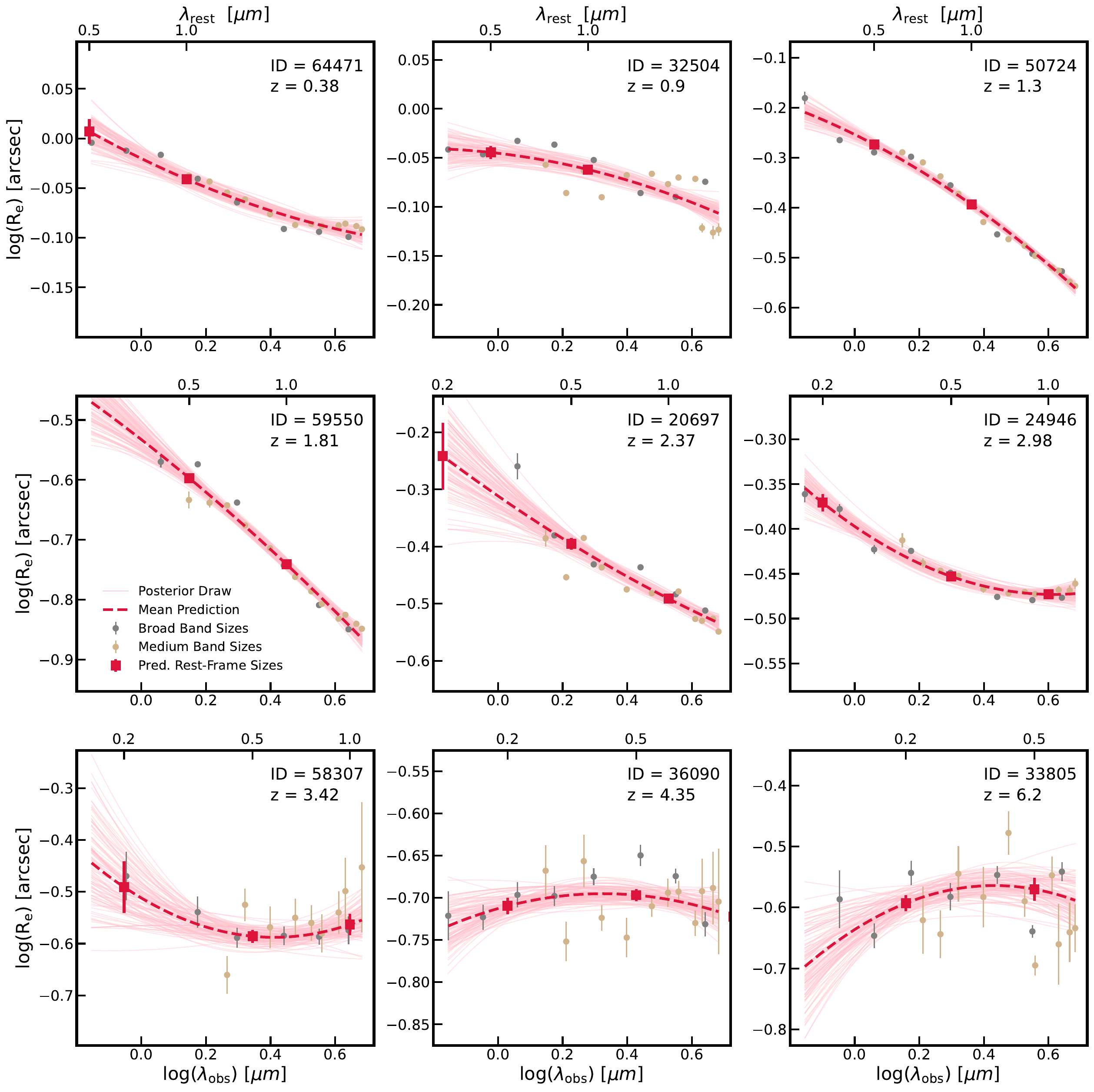}
    \caption{Nine examples at $z\sim 0.3 -8$ whose observed sizes are fitted as a quadratic function of wavelength, from which we derive sizes at several rest-frame wavelengths. We show the observed sizes in brown (medium bands) or grey (broad bands) and rest-frame UV ($0.2 \mu m$), optical ($0.5 \mu m$), and infrared ($1 \mu m$) sizes in red. Pink lines represent the posterior draws from the fit, and the dashed red line represents the best-fitting size as a function of wavelength (corresponding to the mean of the posterior). Overall, our fitting method provides robust estimations and uncertainties of galaxy rest-frame sizes while being resilient to any outliers in the observed sizes. These fits also illustrate the diversity of color gradients in these galaxies.}
    \label{fig:rest-size}
\end{figure*}

In Figure \ref{fig:rest-size}, we showcase nine instances of size interpolation that are uniformly distributed over $0.3<z<8$. The best-fitting individual observed sizes and uncertainties are shown in grey (broad bands) or brown (medium bands). We overlay the best-fitting size as a function of wavelength as a red dashed curve, as well as 100 posterior draws as pink lines. The best-fitting function of size is taken as the median of the posterior draws. We show the inferred rest-frame UV ($\rm 0.2\mu m$), optical ($\rm 0.5\mu m$), and NIR ($\rm 1\mu m$) sizes and uncertainties as red squares with error bars, where applicable. Given the NIRCam filter coverage of observed wavelength ($\rm 0.7-4.8\mu m$), we reliably infer rest UV sizes at $\rm z<1.8$, optical sizes at $\rm 0.1<z<11$, and NIR sizes at $\rm z<5$, as long as the given rest wavelength falls between the effective wavelengths of any two robust observed size measurements or within $\rm 0.1\,dex$ (logarithmically in microns) of an available robust observed size. 

\begin{figure}[!tb]
    \centering
    \includegraphics[width = 1.\linewidth]{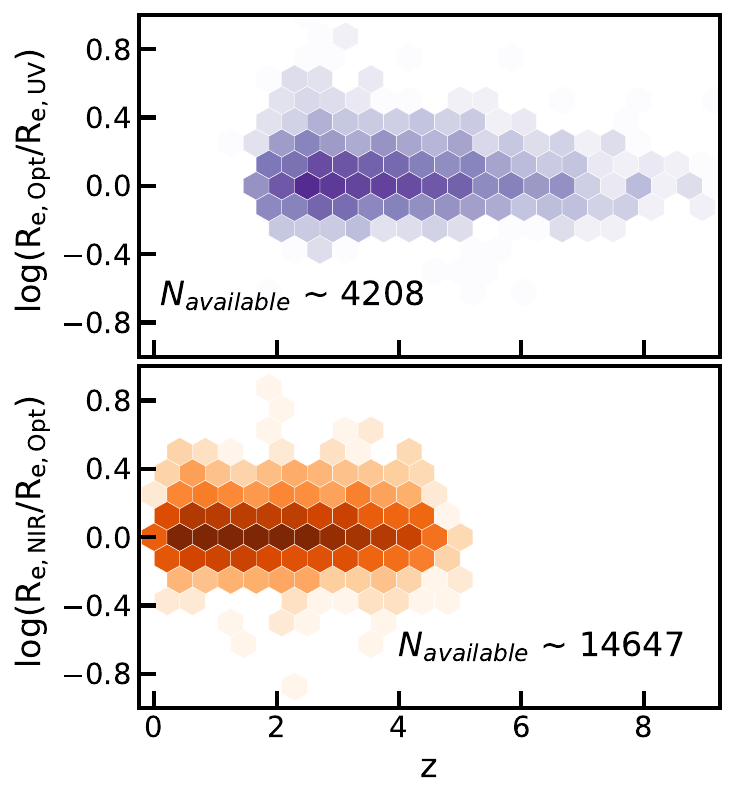}
    \caption{In the top panel, we show the rest UV-optical color gradient strength versus redshift for all sources in this sample where both rest UV and optical sizes are available. In the bottom panel, we similarly show the rest optical-NIR color gradient strength versus redshift. The wide scatter of these distributions indicates the large diversity in color gradient strength among the galaxy population probed by this sample.}
    \label{fig:color-gradient distribution}
\end{figure}

Overall, these fits of $\rm R_{e}(\lambda)$ are resilient to outliers in the observed sizes and provide a robust inference of half-light size and its uncertainties at these rest-frame wavelengths. One caveat is that in a few cases (e.g., ID: 58307), there are systematic differences between sizes measured in broad bands and medium bands. This systematic difference originates from the difference in effective depth between the broad band and the medium band image mosaics. In Figure \ref{fig:rest-size}, the uncertainties in rest-frame size are entirely driven by measurement uncertainties in observed sizes; we do not incorporate redshift uncertainties. The vast majority of sources in this dataset rely on photometric redshifts. The redshift uncertainty can dominate the uncertainty in rest-frame sizes if these sources have a strong color gradient, are located at $z<1.4$ where the angular diameter distance evolves rapidly with redshift, or are positioned in the lensing configuration where the magnification value is sensitive to redshift. The uncertainty in rest frame sizes shown in this figure is underestimated relative to the values we report in the Rest Frame Size catalog, where we do incorporate redshift uncertainties.

These fits of galaxy size as a function of wavelength shed light on the intrinsic color gradients in these galaxies. In Figure \ref{fig:color-gradient distribution}, we demonstrate the distribution of color gradient strength versus redshift for this sample. The color gradient strength is represented by the ratio of half-light radii at rest-frame UV versus optical (or optical and NIR). While the overall median logarithmic color gradient strength is $\sim 0$, the wide range of these indicates the large diversity in color gradient strength among the galaxy population across cosmic time. The significant color gradients among some of these galaxies can be driven by their differential structure of stellar populations or dust geometry, and their strengths can systematically depend on the spectral type \citep[e.g.,][]{Suess.etal.2021}. We defer a concise study of galaxy color gradients using this sample to the future.

\subsection{A First Look at the Rest-Frame Optical Size-Mass Relation over 8 Billion Years of Cosmic History}

These derived physical half-light sizes at various rest-frame wavelengths enable new dimensions in which these galaxy populations behind Abell 2744 can be studied. This is in addition to the wealth of ancillary data, including the stellar population properties from the \texttt{SPS} catalog \citep{Wang.etal.2024}. One excellent example among many scientific opportunities offered by this combined dataset is the redshift-evolving scaling relationship between galaxy stellar mass and optical size for star-forming galaxies versus quiescent galaxies. This scaling relationship imprints valuable information regarding the galaxy growth history throughout cosmic time and places important constraints on cosmological galaxy simulations \citep[e.g.,][]{Furlong.etal.2017,Genel.etal.2018}. In the previous decade, this relation has been successfully characterized from the local universe up to the cosmic noon ($\rm z<3$), from the massive galaxies to dwarf galaxies, through ground-based and/or space-based imaging \citep[e.g.,][]{vanderWel.etal.2014,Carlsten.etal.2021,Mowla.etal.2019,Kawinwanichakij.etal.2021,Nedkova.etal.2021,Cutler.etal.2022}. Robustly characterizing this relation for galaxies at higher redshift ($\rm z>3$) was not possible until the launch of JWST, which provides access to high-resolution NIR imaging out to $\rm \sim 5\mu m$ \citep[e.g.,][]{Ormerod.etal.2024,Allen.etal.2025,Miller.etal.2025}. 

Taking advantage of the lensing cluster, Abell 2744, the UNCOVER/MegaScience dataset now pushes the measurement limit in stellar mass and size to even fainter and smaller galaxies than previous JWST samples from blank legacy fields. The 20-band NIRCam photometry from UNCOVER/MegaScience also enables robust photometric redshift estimation for these galaxies \citep{Suess.etal.2024}, covering the rest-optical for over 8 billion years of cosmic history, up to $z\sim8$. Next, we showcase the unprecedented survey coverage of the galaxy population in terms of stellar mass, optical size, and redshift in this dataset, which undeniably attests its legacy value to the study of galaxy size growth. In this paper, we qualitatively compare the distribution of the star-forming and quiescent populations to previously measured scaling relations, and we defer the updated measured scaling relations to the companion paper \citep{Miller.etal.2026}.

\begin{figure*}[!htb]
    \centering
    \includegraphics[width = 1.\textwidth]{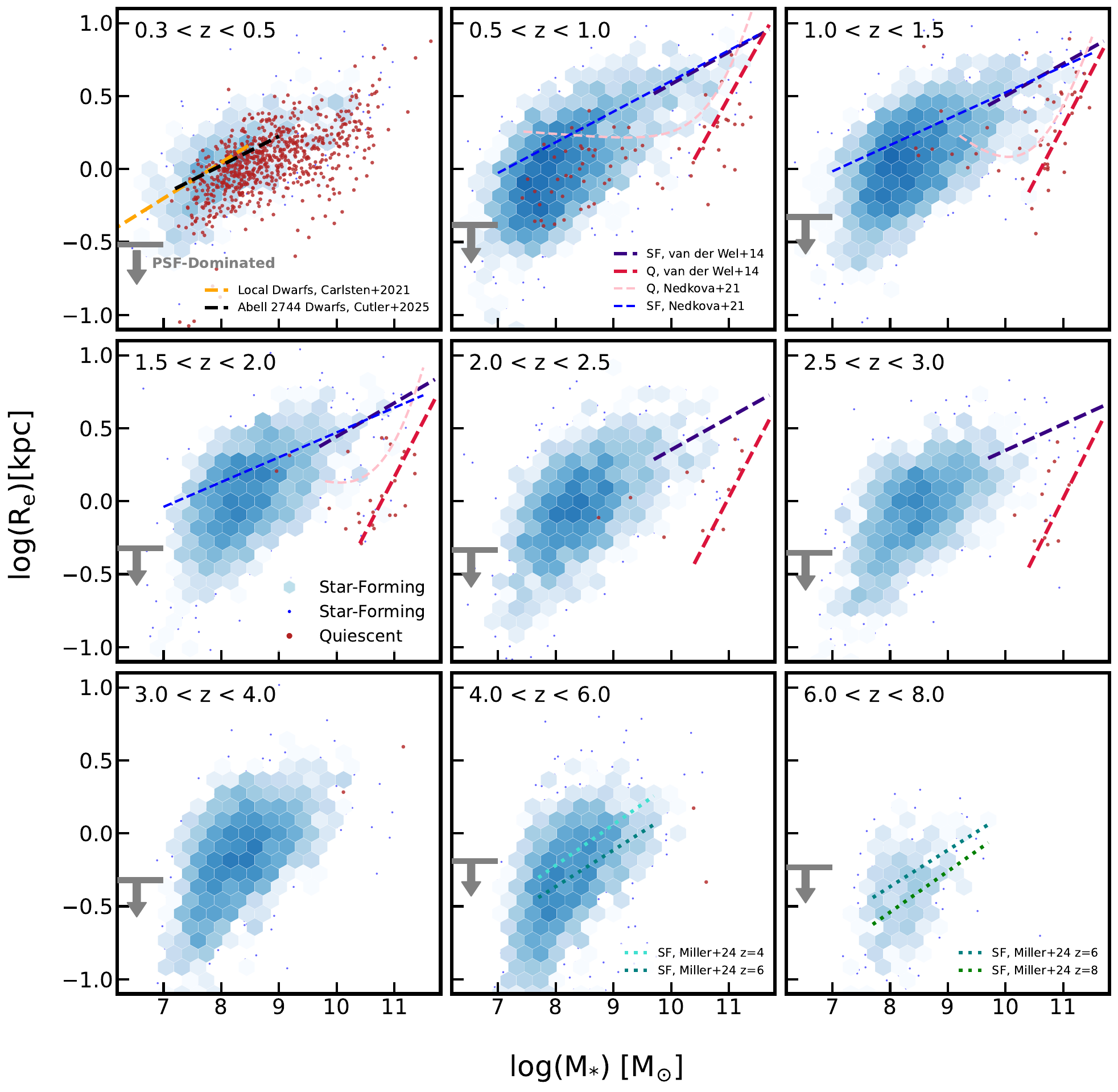}
    \caption{The distribution in mass versus rest-frame optical size for galaxies with available measurements from this work binned by redshift. We show individual quiescent galaxies as red dots. For star-forming galaxies, we show their distributions as 2D histograms with blue hexagon bins. Darker bin color corresponds to higher source density. In bins containing fewer than three star-forming galaxies, we neglect binning and show them as individual blue dots. The star-forming galaxies and quiescent galaxies are separated by rest-frame {\it ugi} color criteria \citep{Antwi-Danso.etal.2023}. We mark the size regime where measured sizes are comparable to PSF widths with gray arrows in each panel. Several literature mass-size relations are plotted for comparison. In general, the mass-size distribution of galaxies in this work follows previous literature relations on the massive end. The galaxies in this work span an unprecedented range in mass, size, and redshift, enabling the characterization of the mass-size relation in a wider parameter space.}
    \label{fig:rf_size_vs_lit}
\end{figure*}

In Figure \ref{fig:rf_size_vs_lit}, we show the rest-frame ($\rm 0.5\mu m$) half-light radii versus stellar mass for star-forming (blue hexagon bins) and quiescent galaxies (red dots) in the UNCOVER/MegaScience catalogs, out to $z=8$. For star-forming galaxies, darker bin color means higher source density; individual blue points are shown for bins with fewer than 3 galaxies. The star-forming and quiescent galaxies are classified through rest-frame synthetic {\it ugi} color criteria \citep{Antwi-Danso.etal.2023}, as recommended in \cite{Wang.etal.2024}. We apply an evolving redshift quality cut:
\begin{equation}
   \rm  z_{50}> (z_{84}-z_{16})/0.6 -1,
\end{equation}
where $z_{N}$ refers to the Nth percentile of the redshift posterior value, to omit any sources with large redshift uncertainties. We also only show the galaxies with moderate magnification ($\rm \mu <4$), excluding highly magnified systems for which the lensing correction is more complex.

For each redshift bin, we have also estimated a rudimentary ``measurement" limit in size (grey arrow), below which the size is dominated by the PSF, estimated as follows. We first take the PSFs from the band that is the closest to rest-frame $\rm 0.5\mu m$ in terms of effective filter wavelength, given the median redshift of each bin. We then take the angular radius at which 50\% of the total energy is enclosed in these PSFs \citep{Weaver.etal.2024}, convert them to a physical size using the bin median redshift, and superficially divide these sizes by a factor of $\sqrt{\mu}$ to account for lensing magnification, where we take $\mu \sim 1.2$ from the median magnification in this sample.

We overlay the mass-size relations reported from various literature in the panels at the corresponding redshifts \citep{vanderWel.etal.2014,Carlsten.etal.2021,Nedkova.etal.2021,Miller.etal.2025}. One caveat is that we have shifted the original mass-size relation from \cite{vanderWel.etal.2014} in stellar mass by $\rm +\,0.2 \,dex$ to account for systematic differences in stellar mass due to star formation history parameterization \citep[e.g., see][]{Leja.etal.2019}. At $z>4$, the overall mass-size distribution of star-forming galaxies in this sample agrees with the mass-size relation reported by \cite{Miller.etal.2025}. \cite{Miller.etal.2025} also selects a similar sample of $\rm 4<z<8$ galaxies from the UNCOVER/MegaScience dataset. A critical difference is that \cite{Miller.etal.2025} requires the S\'ersic parameters to vary smoothly with wavelength, in contrast with our method, which performs the size inference in all bands independently and then fits these sizes to obtain the size as a function of wavelength. The excellent agreement between our rest-frame size distributions and these reported relations speaks to the robust nature of our methodology. This agreement also bolsters the assumption that S\'ersic parameters vary smoothly with wavelength in most galaxies.

At $z<3$, our galaxy sample covers the mass regime probed by previous HST samples ($\rm 10^{7}M_{\odot}<M_{*}<10^{11.5}M_{\odot}$; \citealp{vanderWel.etal.2014,Nedkova.etal.2021}). At face value, the mass-size distributions of both star-forming and quiescent galaxies in this sample largely agree with the relations derived from HST sample in the massive regime ($\rm M_{*}>10^{10}M_{\odot}$). However, these HST-measured relations ($z<2$, \citealp{Nedkova.etal.2021}) apparently predict a much larger median size than this sample at a given mass for dwarf star-forming or quiescent galaxies ($\rm M_{*}<10^{9}M_{\odot}$). It is likely because the sizes of the smallest galaxies in previous HST samples were systematically overestimated, since the angular resolution of those HST imaging ($\sim 0.06''$) is three times worse than JWST/NIRCam ($\sim 0.02''$). Alternatively, the previous HST samples could be incomplete and missing some of the faintest, smallest galaxies, since those HST images were shallower than the NIRCam images in UNCOVER/MegaScience. Intriguingly, the dwarf ($\rm M_{*}<10^{9}M_{\odot}$) quiescent population at $0.5<z<1.5$ seems to have lower median size than that reported in \cite{Nedkova.etal.2021}. We suspect that such a discrepancy originates from the different selection of the quiescent population rather than reflecting an actual disagreement. In addition, it is probable that some of the quiescent dwarfs shown at $0.5<z<1.0$ can be cluster members at lower redshifts or misclassified star-forming dwarfs. Any contamination in the dwarf quiescent galaxy in this demo can arise from large photometric redshift errors or inaccurate rest-frame {\it u-g} color estimation, since these properties are based on NIRCam photometry that has poor rest-frame coverage (missing $\rm \lambda_{rest}<0.35 \mu m$) at low redshifts ($z<1$). A detailed and comprehensive comparison and discussion between these galaxies in UNCOVER/MegaScience and those in \cite{Nedkova.etal.2021} at $1<z<1.5$ has been presented in \cite{Cutler.etal.2024}, though the quiescent dwarfs at $0.5<z<1$ in this sample has not yet been discussed. 

At $z<0.5$, this sample contains $\sim 1500 $ quiescent dwarf galaxies that are cluster members of Abell 2744. This unique sample of cluster dwarf galaxies (down to $\rm M_{*}>10^{7}M_{\odot}$) enables the study of quiescent dwarf structural evolution across cluster environments when combined with local universe samples \citep{Carlsten.etal.2021,Eigenthaler.etal.2018,Cutler.etal.2025}.

At face value, the mass-size distribution of dwarf star-forming galaxies at $z<3$ appears to have a steeper slope than that of massive star-forming galaxies, or that of dwarfs reported by \cite{Nedkova.etal.2021}. However, this apparent disagreement can be a deceptive artifact because we are only plotting these galaxies at their 50th percentile stellar mass value. For these dwarf galaxies, the uncertainties in stellar mass ($\rm \sim 0.22 \,dex$) are significantly larger than those in size ($\rm \sim 0.04 \,dex$). This naturally broadens the horizontal scatter and flattens the relation at lower mass. Furthermore, the uncertainties in stellar mass, size, and redshift are all correlated for these galaxies. Due to gravitational lensing, both stellar mass and size are scaled with the magnification, which depends on the redshift. For the majority of this sample that relies on photometric redshift, the redshift posterior distributions $\rm p(z)$ can be complicated and substantial at low-redshifts where the JWST wavelength coverage is less powerful, which can further propagate into uncertainties in stellar mass and size. To robustly quantify the redshift evolving mass-size relation using this sample and meaningfully compare it to any other studies, one must carefully incorporate these complicated, correlated uncertainties when fitting these data for the median galaxy size as a function of both stellar mass and redshift. We defer such a detailed analysis to a companion study that will be released concurrently with this work \citep{Miller.etal.2026}.

\section{Summary}\label{sec: summary}

We present and publicly release single-component S\'ersic fits of 28,274 sources from the UNCOVER/MegaScience catalogs using the 20 band JWST/NIRCam image mosaics in Abell 2744. The best-fitting S\'ersic parameters and their corresponding uncertainties are inferred using \texttt{pysersic}, a Bayesian structural fitting tool. For any sources with $\rm SNR>10$, we estimate the S\'ersic parameter posterior with the stochastic variational inference using a multivariate normal method. We also perform a second parameter posterior estimation for sources with $\rm 10<SNR<100$, using MCMC sampling. We conduct a thorough quality check for these fits and classify them into three levels of quality: great quality, robust quality, and failed instances. All raw results from these fits and corresponding quality information are presented in the \texttt{RAW} catalog. We additionally produce two \texttt{SUPER} catalogs that include only great quality, or great and robust quality. These \texttt{SUPER} catalogs also opt to include results from \texttt{MCMC} if available, or \texttt{SVI-MVN} if otherwise. All sizes reported in the \texttt{RAW}, \texttt{SUPER}, and \texttt{SUPER\_DUPER} catalogs are not corrected for lensing magnification. 

To explore the systematic uncertainties in these fits, we perform recovery tests on simulated galaxy cutouts in select bands, which are made with real empty sky cutouts and simulated 2D S\'ersic profiles. We confirm these S\'ersic parameters have fractional systematic uncertainties less than $5\%$ in the SNR regime of this catalog ($\rm SNR>10)$. Based on these results, we recommend that users inflate the random uncertainties reported by \texttt{pysersic} by a factor of 2 for flux and half-light radius (or a factor of 1.5 for S\'ersic index) to compensate for their slight underestimation. 

For each galaxy with robust sizes in at least two bands and a robust redshift, we fit the observed size as a function of wavelength. These parameterized fits allow us to extract rest-frame galaxy sizes, which mitigates the bias in size due to the intrinsic color gradient when relying on the fixed effective wavelength of a specific filter. This enables fair comparison of galaxy sizes across redshifts. These fits are presented in the Rest Frame Size catalog. 

Combining these fits with the redshift and stellar mass presented in the \texttt{SPS} catalog, we show that this set of morphological catalogs will enable measurements of the galaxy mass-size relation over an unprecedentedly wide parameter space of redshift ($0.3<z<8$), stellar mass ($\rm 10^{7}\, M_{\odot}<M_{*} <10^{11.5}\, M_{\odot}$), and rest-frame optical size ($\rm 100 \,pc<R_{e}<10\,kpc$), among many other potential topics these fits can unlock. We note that a companion paper will characterize the galaxy mass-size relation in this parameter space \citep{Miller.etal.2026}, leveraging the catalogs produced by this work. We release these catalogs to the community online at \url{https://zenodo.org/records/18808251}.

\section{Acknowledgments}
This work is based in part on observations made with the NASA/ESA/CSA James Webb Space Telescope. The data were obtained from the Mikulski Archive for Space Telescopes at the Space Telescope Science Institute, which is operated by the Association of Universities for Research in Astronomy, Inc. (AURA), under NASA contract NAS 5-03127 for JWST. These observations are associated with JWST-GO-2561, JWST-GO-4111, JWST-GO-2883, JWST-GO-3516, JWST-3538, JWST-ERS-1324, and JWST-DD-2756. Support for program JWST-GO-2561 was provided by NASA through a grant from the Space Telescope Science Institute under NASA contract NAS 526555. The specific original observations used to produce our UNCOVER catalogs can be accessed via \dataset[doi: 10.17909/7yvw-xn77]{http://dx.doi.org/10.17909/7yvw-xn77}.

Some of the data products presented herein were retrieved from the Dawn JWST Archive (DJA). DJA is an initiative of the Cosmic Dawn Center (DAWN), which is funded by the Danish National Research Foundation under grant DNRF140.

The Cosmic Dawn Center is funded by the Danish National Research Foundation (DNRF) under grant \#140. 

RB gratefully acknowledges support from the Research Corporation for Scientific Advancement (RCSA) Cottrell Scholar Award ID No: 27587.

P. Dayal warmly acknowledges support from an NSERC discovery grant (RGPIN-2025-06182).

The work of CCW is supported by NOIRLab, which is managed by the Association of Universities for Research in Astronomy (AURA) under a cooperative agreement with the National Science Foundation.

This research was supported in part by the University of Pittsburgh Center for Research Computing and Data, RRID:SCR\_022735, through the resources provided. Specifically, this work used the H2P cluster, which is supported by NSF award number OAC-2117681. 

DJS and JRW acknowledge that support for this work was provided by The Brinson Foundation through a Brinson Prize Fellowship grant. 

TBM was supported by a CIERA Fellowship. This work used computing resources provided by Northwestern University and the Center for Interdisciplinary Exploration and Research in Astrophysics (CIERA). This research was supported in part through the computational resources and staff contributions provided for the Quest high performance computing facility at Northwestern University, which is jointly supported by the Office of the Provost, the Office for Research, and Northwestern University Information Technology.

\facilities{ 
JWST (NIRCam)
}

\software{Astropy \citep{astropy:2013,astropy:2018,astropy:2022}, Scipy \citep{2020SciPy}, Photutils \citep{photutils}, Numpy \citep{Van.Der.Walt.etal.2011}, Matplotlib \citep{Hunter.2007}, pysersic \citep{Pasha_and_Miller.2023}}

\appendix

\section{Catalog Completeness }\label{appendix:A}
\begin{figure*}[!ht]
    \centering
    \includegraphics[width = 1.\textwidth]{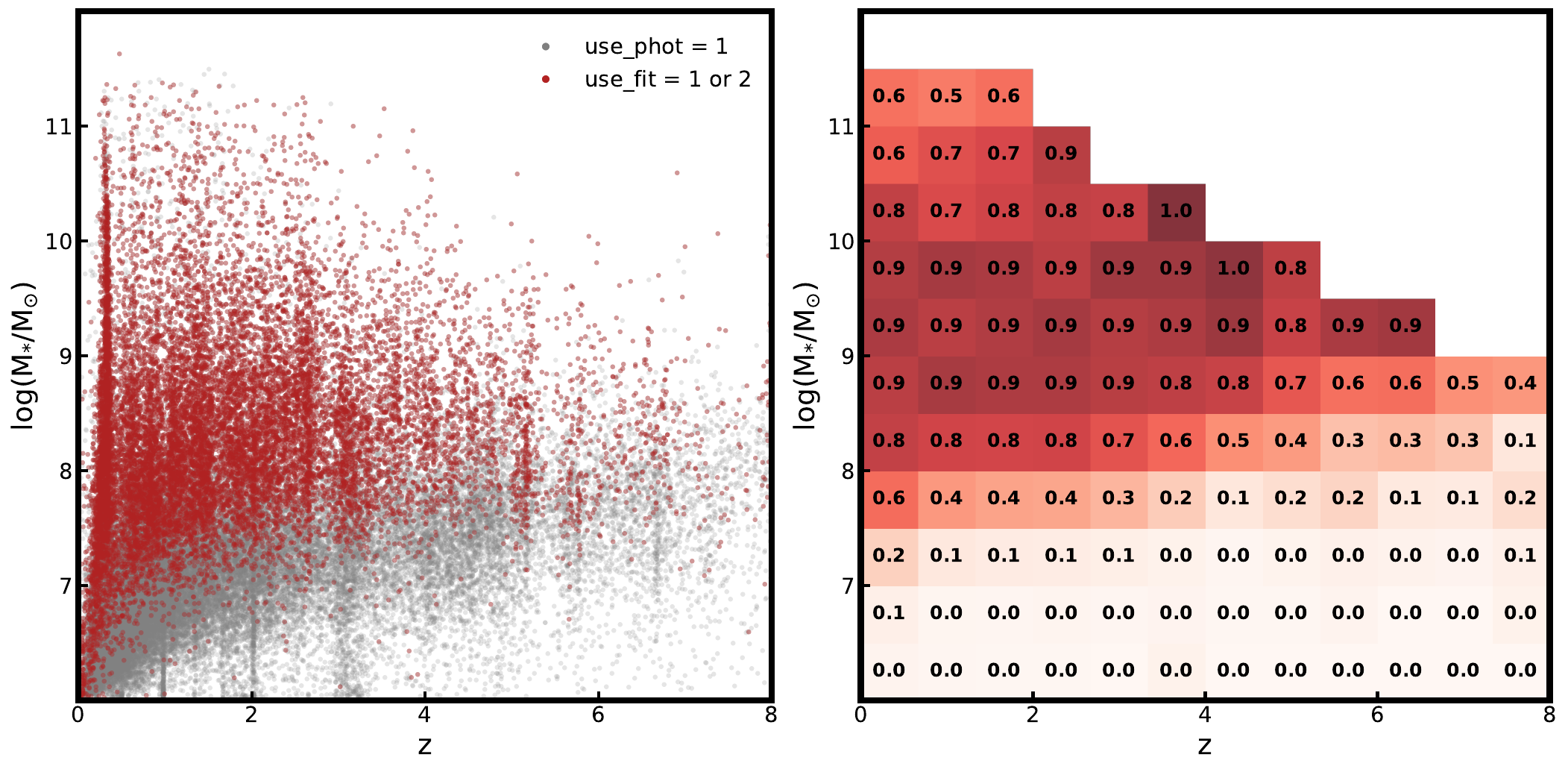}
    \caption{The left panel shows the redshift versus stellar mass for valid sources in F444W in the UNCOVER/MegaScience photometric catalog (\texttt{use\_phot = 1}; grey) and all sources with a great or robust fitin F444W from this work (\texttt{use\_fit = 1 or 2}; red). The right panel shows the net completeness of our fits in F444W, binned by redshift and stellar mass. We define completeness as the fraction of \texttt{use\_fit = 1 or 2} sources among \texttt{use\_phot = 1} sources. We omit any bins containing fewer than 10 \texttt{use\_phot = 1} sources. Overall, we achieve $> 80\%$ net completeness at $\rm M_{*}>10^{8}M_{\odot}$ and $z<2$; $\rm M_{*}>10^{8.5}M_{\odot}$ and $2<z<4$; $\rm M_{*}>10^{9}M_{\odot}$ and $4<z$. }
    \label{fig:catalog completeness}
\end{figure*}

Due to the global $\rm SNR>10$ target selection and various quality evaluation criteria listed in Section \ref{sec: flags}, the resulted sample of great or robust (\texttt{use\_fit = 1 or 2}) structural measurements are incomplete relative to all the available sources in the UNCOVER/MegaScience survey volume. The completeness of these fits likely has complicated dependency on the surface brightness distribution of surveyed sources, which correlates to a diverse number of source properties, such as stellar mass, redshift, morphology, dust, age, etc. In this section, we focus on evaluating the completeness of fits in this work as a function of stellar mass and redshift. We define the completeness as the fraction of sources with great or robust fit (\texttt{use\_fit = 1 or 2}) among total available photometric sources (\texttt{use\_phot = 1}). We opt to only present this diagnostic in F444W. However, we note this diagnostic can be easilly repliacted or extended to other bands, given the UNCOVER/MegaScience \texttt{SPS} catalog, photometric catalog, and the RAW catalog from this work. In Figure \ref{fig:catalog completeness}, we show the scatter of \texttt{use\_fit = 1 or 2} sources and \texttt{use\_phot = 1} sources in stellar mass and redshift in the left panel. In the right panel, we show the completeness (the fraction of these two sources) binned by stellar mass and redshift. In general, this work achieve $> 80\%$ completeness above $\rm 10^{8}M_{\odot}$ at $z<2$, $\rm 10^{8.5}M_{\odot}$ at $2<z<4$, and $\rm 10^{9}M_{\odot}$ at $4<z$. We note that the completeness drops to $\sim 60\%$ at $\rm M_{*}>10^{10.5}M_{\odot}$ and $z<2$, where available sources are relatively rare. These sources are the brightest among the entire sample and the incompleteness in this regime is entirely due to fits dropped out by our quality evaluation. Among these dropouts, roughly half of the sources have bright asymmetries such that their $\chi_{pp}^{2}$ exceeds the hard limit we enforce, which is already generously tolerate towards bright sources. The remaining dropout sources have unmatched parameter posteriors and MAP values, and they are determined as \texttt{use\_fit = 0} by criterion B). This type of issues likely originates from implicit bugs in the implementation of \texttt{pysersic}. We defer a thorough investigation of these issues to the future and will update the released catalogs once these issues are resolved.

\section{Sub-pixel Shift in Centroid Positions per Filter }\label{appendix:B}

\begin{figure*}[!ht]
    \centering
    \includegraphics[width = 1.\textwidth]{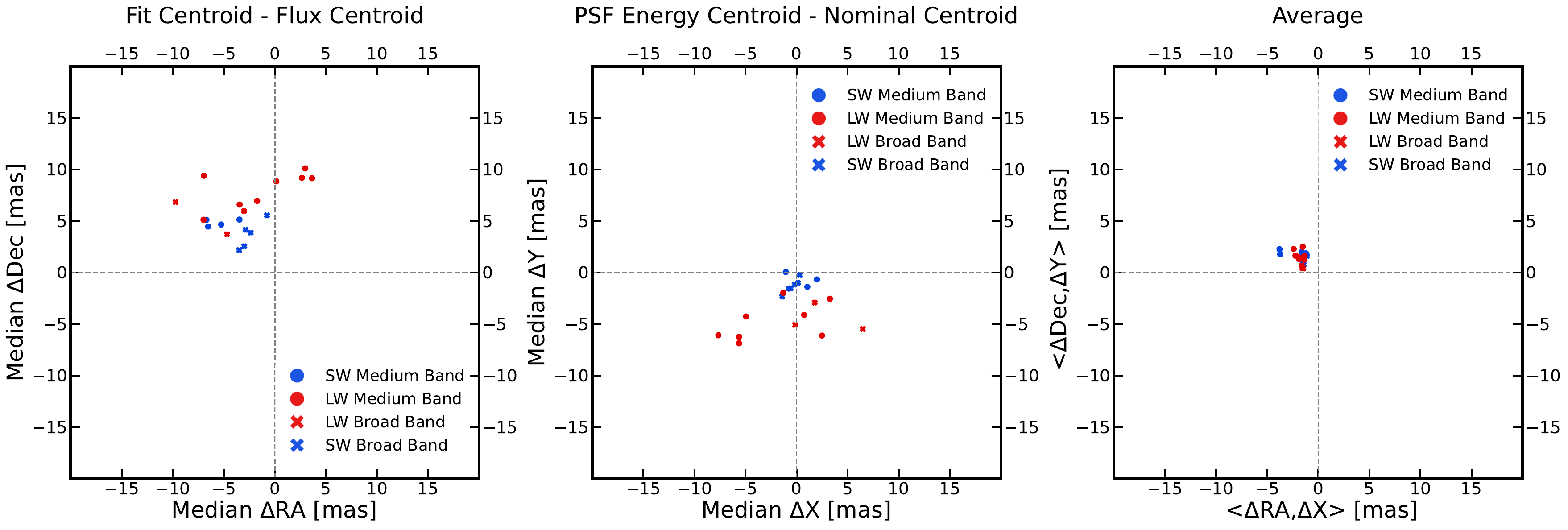}
    \caption{The left panel shows the median difference between the best-fitting S\'ersic centroids and the flux centroids in each band in mas. The middle panel shows the difference between the PSF energy centroids and the PSF nominal centroids in each band in mas. The right panel shows the average of the centroid offsets shown in the left and middle panels. It is likely the systematic sub-pixel centroid offsets in our fits are driven by the centroid offsets in PSFs, since the latter have similar amplitudes but opposite directions when compared to the former.}
    \label{fig:sysmatic shift}
\end{figure*}

For source fits classified as great (\texttt{use\_fit = 2}) in this work, the centroids of the best-fitting 2D S\'ersic systematically deviate from the flux centroids reported in the photometric catalog. The amplitude of such a systematic offset is extremely minor (much less than the size of a pixel) and varies from filter to filter. In the left panel of Figure \ref{fig:sysmatic shift}, we demonstrate these offsets in all 20 NIRCam filters by showing the median differences between fit centroids and flux centroids in right ascension and declination, in units of milliarcseconds (mas). The amplitude of the offset clearly displays a dependency on the pixel scale, as the offsets in LW bands are typically twice as much as those in SW bands. It is likely that these offsets are driven by the effective centroid of the PSF profiles being unaligned with the nominal geometrical center of the PSF images.

In the middle panel of Figure \ref{fig:sysmatic shift}, we show the offsets of the PSF energy centroids from their nominal centroids in all 20 NIRCam filters. We approximately determine these PSF energy centroids by fitting a single-component 2D Gaussian profile to the PSF profile and adopting the center of the best-fitting Gaussian. The offsets of the PSF energy centroids show similar amplitudes as the offsets of the S\'ersic fit centroids, in roughly the opposite direction. We further confirm this trend by taking the averages of these offsets in each band, which we show in the right panel, and we find these averages to be much closer to zero. There are still systematic offsets ($\rm \sim 2 \, mas$) remaining after accounting for the PSF centroid offsets, which we deem negligible since they are an order of magnitude smaller than a NIRCam pixel in SW.

\section{Catalog Column Descriptions}\label{appendix:C}

With this paper, we release our direct and value-added structural measurements to the community in four catalogs: a \texttt{RAW} catalog containing all original outputs of structural fits and their relevant statistics for quality flagging, a \texttt{SUPER} catalog containing only structural fit results with great or robust quality (\texttt{use\_fit = 2 or 1}, see Section \ref{sec: flags} for details), a aggressively filtered \texttt{SUPER-DUPER} catalog containing only structural fit results with great quality (\texttt{use\_fit = 2}), and a Rest Frame Size catalog containing the parameterized galaxy size as a function of wavelengths and inferred rest-frame sizes. We enumerate the \texttt{RAW} catalog columns in Table \ref{tab:RAW catalog}, the shared columns for \texttt{SUPER} or \texttt{SUPER-DUPER} catalogs in Table \ref{tab:SUPER catalog}, and the columns for the Rest Frame Size catalog in Table \ref{tab:RF catalog}. We note that the half-light radii in \texttt{RAW}, \texttt{SUPER}, \texttt{SUPER-DUPER}, and the observed rest-frame half-light radii are not corrected for lensing magnification and reported in units of arcseconds. The physical rest frame half-light radii reported in the Rest Frame Size catalog are, however, corrected for lensing magnification and have units of kiloparsecs.

\begingroup
\setlength{\tabcolsep}{20pt}
\begin{table*}[t!]
\caption{\texttt{RAW} Catalog Columns}
    \begin{tabular}{ll}
    \hline
    \hline
    Column name & Description \\
    \hline
    id & Object identification number, same as the DR3 photometry catalog\\
    X\_M\_xc/yc\_16/50/84 & Image-frame S\'ersic X/Y centroid posterior percentiles (pixels), in filter X with method M\\
    X\_M\_flux\_16/50/84 & Flux posterior percentiles ($\rm 10nJy$), in filter X with method M\\
    X\_M\_r\_eff\_16/50/84 & Half-light radius posterior percentiles (pixels), in filter X with method M\\
    X\_M\_n\_16/50/84 & S\'ersic index posterior percentiles, in filter X with method M\\
    X\_M\_ellip\_16/50/84 & Ellipcity posterior percentiles, in filter X with method M\\
    X\_M\_theta\_16/50/84 & Angular position posterior percentiles, in filter X with method M\\
    X\_M\_frac\_masked & Fraction of the pixels masked within 4 half-light radii, in filter X with method M\\
    X\_flux\_MAP & Flux at maximum a posteriori, in filter X\\
    X\_r\_eff\_MAP & Half-light radius at maximum a posteriori, in filter X\\
    X\_n\_MAP & S\'ersic index at maximum a posteriori, in filter X\\
    X\_chi2\_pp & Chi squared per pixel \\
    X\_ess\_b\_min  & Minimum bulk effective sample size, if \texttt{MCMC} \\
    X\_ess\_t\_min  & Minimum tail effective sample size, if \texttt{MCMC} \\
    X\_rhat\_max  & Maximum Rhat convergence diagnostic, if \texttt{MCMC} \\
    X\_M\_use\_fit  & Quality flags of the fit, in filter X with method M (see Section \ref{sec: flags})\\
    mu\_flag & Weak or strong magnification flag (= 0, if $\mu < 4$; =1 , if $\mu>4$)\\
    flag\_nearbcg & 1, if within $\rm 3''$ of a known bCG, same as the DR3 photometry catalog\\ 
      \hline
    \end{tabular}
    \tablecomments{X = filter name, as listed in Section \ref{sec: data}. M = method name, \texttt{MCMC} or \texttt{SVI-MVN}, as described in Section \ref{sec: implementation}.}
    \label{tab:RAW catalog}
\end{table*}
\endgroup

\begingroup
\setlength{\tabcolsep}{20pt}
\begin{table*}[t!]
\caption{\texttt{SUPER}/ \texttt{SUPER DUPER} Catalog Columns}
    \begin{tabular}{ll}
    \hline
    \hline
    Column name & Description \\
    \hline
    id & Object identification number, same as the DR3 photometry catalog \\
    RA & R.A. J2000 (degrees), same as the DR3 photometry catalog \\
    Dec & Decl. J2000 (degrees), same as the DR3 photometry catalog \\
    X\_flux\_16/50/84 & Flux posterior percentiles, in filter X\\
    X\_r\_eff\_16/50/84 & Half-light radius posterior percentiles, in filter X\\
    X\_n\_16/50/84 & S\'ersic index posterior percentiles, in filter X\\
    X\_ellip\_16/50/84 & Ellipcity posterior percentiles, in filter X\\
    X\_dist2phot\_cen & S\'ersic centroid's distance to the photometric catalog centroid, in arcseconds\\ 
    mu\_flag & Weak or strong magnification flag (same as \texttt{RAW})\\
    flag\_nearbcg & 1, if within $\rm 3''$ of a known bCG, same as the photometry catalog DR3\\ 
      \hline
    \end{tabular}
    \tablecomments{X = filter name, as listed in Section \ref{sec: data}. For these ``distilled'' catalogs, we prioritize choosing the parameter posterior percentiles obtained from \texttt{MCMC} sampling, when reliable outputs are available (see Section \ref{sec: catalogs} for details). Otherwise, we choose the parameter posterior percentiles obtained from \texttt{SVI-MVN} approximation. }
    \label{tab:SUPER catalog}
\end{table*}
\endgroup

\begingroup
\setlength{\tabcolsep}{20pt}
\begin{table*}[t!]
\caption{Rest Frame Size Catalog Columns}
    \begin{tabular}{ll}
    \hline
    \hline
    Column name & Description \\
    \hline
    id & Object identification number, same as the DR3 photometry catalog\\
    N\_filt & Number of filters used in the fitting of size as a function of wavelength\\
    loglam\_min & Shortest wavelength filter, in units of $\log(\lambda/\mu m)$ used in the fitting process\\
    loglam\_max & Longest wavelength filter, in units of $\log(\lambda/\mu m)$ used in the fitting process\\
    pX\_16/50/84 & Posterior percentiles for the parameters of the polynomial in Eqn.~\ref{eqn:size_lambda}\\
    r\_eff\_Y\_as\_16/50/84 & Posterior percentiles for the angular size, in arcseconds. \\
    r\_eff\_Y\_kpc\_16/50/84 & Posterior percentiles for the physical size, in kpc.\\
    \hline
    \end{tabular}
    \tablecomments{X corresponds to 0,1,2, the coefficients of the polynomial defining size as a function of wavelengths. Y corresponds to one of three rest frame wavelengths: uv - $0.2\ \mu m$, opt - $0.5\ \mu m$, and nir - $1\ \mu m$. This catalog is based on measurements from the \texttt{SUPER} catalog described above and only includes galaxies that have at least two filters in this catalog and the wavelength dependence is successfully fit with a polynomial.}
    \label{tab:RF catalog}
\end{table*}
\endgroup

\section{Recovery Test in F200W and F250M }\label{appendix:D}

\begin{figure*}[!ht]
    \centering
    \includegraphics[width = 1.\textwidth]{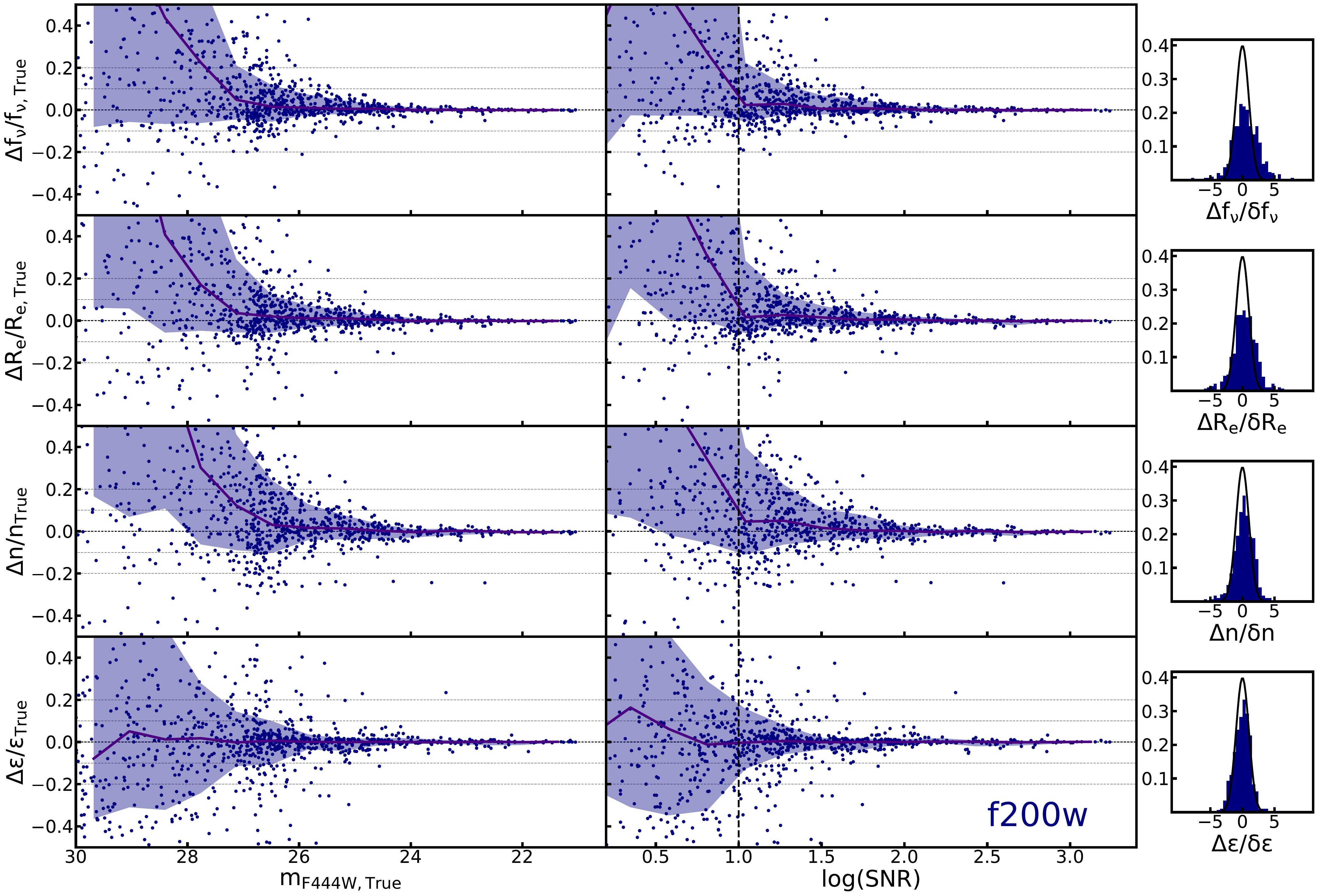}
    \caption{Recovery test in F200W: The plotting convention follows Figure \ref{fig:recovery_f444w}. In F200W, \texttt{pysersic} retrieves S\'ersic parameters within fractional errors of 10\% for the majority ($68\%$) of the instances at a given magnitude or SNR, down to $\rm SNR\sim 10$. Similar to the recovery test in F444W, \texttt{pysersic} systematically underestimates the uncertainty for flux, half-light radius, and S\'ersic index by a factor of roughly 2 (or 1.5, in the case of S\'ersic index).}
\label{fig:recovery_f200w}
\end{figure*}

\begin{figure*}[!ht]
    \centering
    \includegraphics[width = 1.\textwidth]{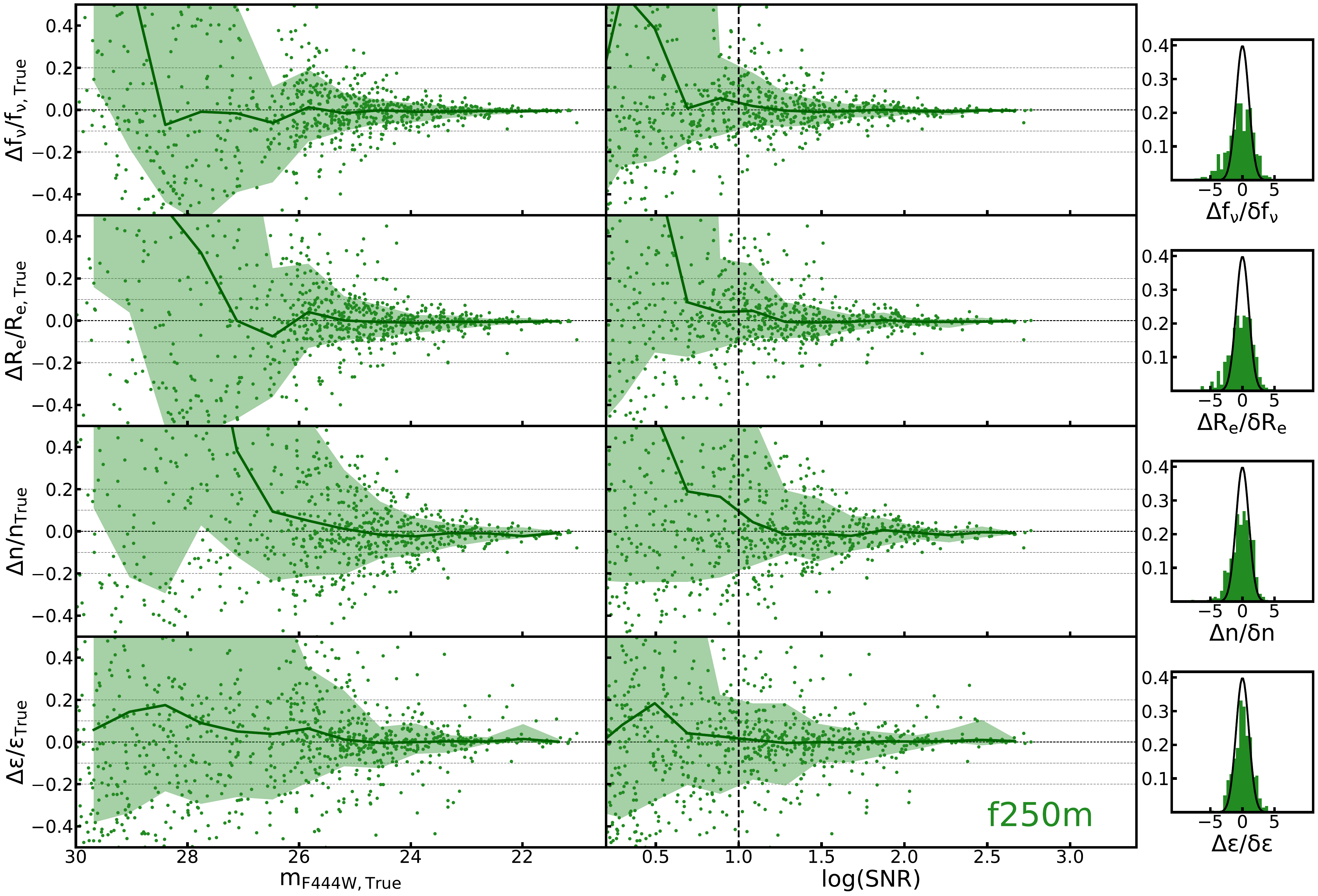}
    \caption{Recovery test in F250M: The plotting convention follows Figure \ref{fig:recovery_f444w}. In F250M, \texttt{pysersic} retrieves S\'ersic parameters within fractional errors of 10\% for the majority ($68\%$) of the instances at a given magnitude or SNR, down to $\rm SNR\sim 20$. Similar to the recovery test in F444W, \texttt{pysersic} systematically underestimates the uncertainty for flux, half-light radius, and S\'ersic index by a factor of roughly 2 (or 1.5, in the case of S\'ersic index).}
\label{fig:recovery_f250m}
\end{figure*}

In Figure \ref{fig:recovery_f200w} and \ref{fig:recovery_f250m}, we present the remaining results of our injection-recovery tests in F200W and F250M, following the same format of the test in F444W, which is discussed in detail in Section \ref{sec: recovery test}. Overall, we find similar trends in \texttt{pysersic} retrieved parameters in these bands as those in F444W: there is little ($\rm <2\%$) or almost no systematic offset from 0 in the median of these fractional errors at $\rm SNR>10$. In F200W, this SNR limit approximately corresponds to a magnitude of 27, while in shallower F250M, it corresponds to a magnitude of 26. By comparing the distribution of fractional errors in these parameters at $\rm SNR>10$ to a unit Gaussian distribution, we find the \texttt{pysersic} reported uncertainties to be underestimated by a factor of 2 for flux or half-light radius (or a factor of 1.5 for S\'ersic index). 

\bibliography{sample631}{}
\bibliographystyle{aasjournal}

\end{document}